\documentclass[11pt,preprint]{aastex}

\newcommand{\tsph}{TreeSPH}
\newcommand{\lya}{Ly$\alpha$\ }

\def\ie{i.e.\ }

\def\cf{cf.\ }
\def\be{\begin{equation}}
\def\ee{\end{equation}}
\def\bdm{\begin{displaymath}}
\def\edm{\end{displaymath}}
\def\ea{et al.}

\def\kmpersec{km s$^{-1}$}
\def\kms{{\rm km}\;{\rm s}^{-1}}
\def\NHI{N_{\rm HI}}
\def\dla{DLA}
\def\lyl{LL}

\def\lta{\mathrel{\rlap{\lower 3pt\hbox{$\mathchar"218$}}
    \raise 2.0pt\hbox{$\mathchar"13C$}}}
\def\hinv{$h^{-1}$}

\newbox\grsign \setbox\grsign=\hbox{$>$} \newdimen\grdimen \grdimen=\ht\grsign
\newbox\simlessbox \newbox\simgreatbox
\setbox\simgreatbox=\hbox{\raise.5ex\hbox{$>$}\llap
     {\lower.5ex\hbox{$\sim$}}}\ht1=\grdimen\dp1=0pt
\setbox\simlessbox=\hbox{\raise.5ex\hbox{$<$}\llap
     {\lower.5ex\hbox{$\sim$}}}\ht2=\grdimen\dp2=0pt
\newcommand{\simgt}{\mathrel{\copy\simgreatbox}}
\newcommand{\simlt}{\mathrel{\copy\simlessbox}}

\newcommand\cdunits{{\rm cm}^{-2}}
\newcommand\vunits{{\rm km}\;{\rm s}^{-1}}
\newcommand\cm{{\rm cm}}
\newcommand\hkpc{h^{-1}\;{\rm kpc}}

\newcommand\hubunits{{\rm km}\;{\rm s}^{-1}\;{\rm Mpc}^{-1}}

\begin{document}

\title{SIMULATIONS OF DAMPED LYMAN-ALPHA AND LYMAN LIMIT ABSORBERS IN
DIFFERENT COSMOLOGIES: IMPLICATIONS FOR STRUCTURE FORMATION AT HIGH
REDSHIFT}

\author{Jeffrey P. Gardner$^{1,2,3}$, Neal Katz$^{4}$, Lars Hernquist$^{5}$,  David H. Weinberg$^6$}
\affil{E-mail:  gardner@phyast.pitt.edu, nsk@kaka.phast.umass.edu,
 lars@cfa.harvard.edu, dhw@astronomy.ohio-state.edu}
\footnotetext[1]
{Department of Physics and Astronomy, University of Pittsburgh,
Pittsburgh, PA 15260}
\footnotetext[2]
{Institute of Astronomy, Madingley Road, Cambridge, CB3 0HA, UK}
\footnotetext[3]
{NSF-NATO Postdoctoral Fellow}
\footnotetext[4]{Department of Physics and Astronomy, University of Massachusetts, Amherst, MA 01003-4525}
\footnotetext[5]
{Department of Astronomy, Harvard University, Cambridge, MA 02138}
\footnotetext[6]
{Ohio State University, Department of Astronomy, Columbus, OH 43210}

\begin{abstract}

We use hydrodynamic cosmological simulations to study damped
Ly$\alpha$ (\dla) and Lyman limit (\lyl) absorption at redshifts
$z=2-4$ in five variants of the cold dark matter scenario:
COBE-normalized (CCDM), cluster-normalized (SCDM), and tilted
($n=0.8$) $\Omega_m=1$ models; and open (OCDM) and flat (LCDM)
$\Omega_m=0.4$ models.  Our standard simulations resolve the formation
of dense concentrations of neutral gas in halos with circular velocity
$v_c \geq v_{c,res} \approx 140\;\kms$ for $\Omega_m=1$ and $90\;\kms$
for $\Omega_m=0.4$, at $z=2$; an additional LCDM simulation resolves
halos down to $v_{c,res} \approx 50\;\kms$ at $z=3$.  We find a clear
relation between HI column density and projected distance to the
center of the nearest galaxy, with \dla\ absorption usually confined
to galactocentric radii less than $10-15$ kpc and \lyl\ absorption
arising out to projected separations of 30 kpc or more.  If we
consider only absorption in the halos resolved by our standard
simulations, then all five models fall short of reproducing the
observed abundance of \dla\ and \lyl\ systems at these redshifts.  To
estimate the absorption from lower mass halos, we fit a power-law to
the relation between absorption area $\alpha$ and halo circular
velocity $v_c$ in our simulations and extrapolate using the Jenkins et
al. (2001) halo mass function; we do not apply this method to the
TCDM model because it has too few halos at the level resolved by our 
simulation.
In the two LCDM simulations, for
which \dla\ results agree well in the mass regime of overlap, the mean
cross-section for \dla\ absorption is $\alpha \approx \pi (0.3
R_{vir})^2$, much larger than the simple estimate $\alpha \sim
\pi(0.1R_{vir})^2$ based on collapse of the baryons to a centrifugally
supported disk ($R_{vir}$ is the halo virial radius).  
The cross sections for \lyl\ absorption are
$\alpha \approx \pi (0.6 R_{vir})^2$, with a dependence on numerical
resolution at the $\sim 25\%$ level.
Detailed examination provides
further evidence of non-equilibrium effects on absorption
cross-section: for example, individual absorbers can be slightly
smaller in more massive halos because gas sinks deeper into the
potential wells, but more massive halos nonetheless have larger
average cross-sections because they are more likely to have multiple
gas concentrations.  Our extrapolation procedure implies that all four
models are consistent with the observed abundance of \dla\ systems if
the fitted $\alpha(v_c)$ extends to $v_c \approx 50-80\;\kms$, and
they may produce too much absorption if the relation continues to $v_c
\la 40\; \kms$.  Matching the observed abundance of \lyl\ systems
requires absorption in halos down to $v_c \approx 30-50\;\kms$.  Our
results suggest that \lyl\ absorption is closely akin to \dla\
absorption, arising in less massive halos or at larger galactocentric
radii but not caused by processes acting on a radically different mass
scale.  Robust tests of cosmological models against the observed
amount of high column density absorption will require simulations of
representative volumes that resolve halos at the low-mass limit where
they cease to harbor high column density absorbers, $30 \simlt v_c
\simlt 60\;\kms$.

\end{abstract}


\section{Introduction}

Systems producing absorption in the spectra of distant quasars offer
an excellent probe of the early Universe.  At high redshifts, they
easily outnumber other observed tracers of cosmic structure, including
both normal and active galaxies.
The interpretation of low column density quasar absorption systems has
undergone somewhat of
a revolution during the past several years, with the recognition that
they may consist of gas aggregating into mildly nonlinear structures
analogous in their dynamical structure to today's galaxy superclusters
\citep{CM94,PM95,ZA95,ZA97,HK96,MC96,bi97,hui97}.
However, damped \lya\ (\dla) absorbers, with neutral hydrogen column densities
$\NHI \geq 10^{20.3}\;\cm^{-2}$, are usually thought to be associated
with the dense interstellar gas of high-redshift galaxies, based on
several lines of circumstantial evidence:
similarity between the column densities of damped systems and the
column densities through typical spiral disks today,
rough agreement between the total mass of atomic hydrogen in
damped absorbers at $z \sim 3$ and the total mass of stars
today \citep{WP98}, measurements of radial extents
$\ga 10\hkpc$ in two \dla\ systems \citep{briggs89,wolfe93},
and direct imaging of a number of \dla\ hosts from ground based
and HST observations \citep{RT98,TR00,djorgovski96,fontana96,moller98,lebrun97}.
The nature of Lyman limit (\lyl) absorbers, with
$\NHI \geq 10^{17.2}\;\cm^{-2}$, is less well understood, though
most models associate them with the outer regions of galaxies
(e.g.\ \citet{mom96}).

Analytic studies based on the Press-Schechter (1974) formalism
suggested that the abundance of \dla\ systems might be a strong
test of cosmological models, potentially ruling out those models
with little power on galaxy scales at $z=3$ \citep{kauffmann94,mo94}.
The most sophisticated of these calculations,
that of Kauffmann (1996), implied that the ``standard'' cold dark
matter model (SCDM, with $\Omega_m=1$, $h\equiv H_0/100\;\hubunits =0.5$,
and a power spectrum normalization $\sigma_8 \approx 0.7$) could
account for the observed abundance of high-redshift DLA systems,
with about 30\% of the absorption at $z=2.5$ occurring in galaxies
with halo circular velocities $v_c > 100\;\vunits$.
Katz et al.\ (1996, hereafter KWHM) presented the first predictions
of the amount of \dla\ and \lyl\ absorption based on 3-dimensional
hydrodynamic simulations, concluding that these simulations of the SCDM model
came within a factor of two of matching the observed \dla\ abundance
but fell nearly an order of magnitude short of reproducing observed \lyl\
absorption.  Ma et al.\ (1997) ``calibrated'' \dla\ estimates from
collisionless N-body simulations against the KWHM SCDM simulations,
then applied this calibration to N-body simulations of cold+hot
dark matter (CHDM) models.  They concluded that the CHDM scenario
failed to reproduce the observed \dla\ abundance even with a neutrino
fraction as low as $\Omega_\nu=0.2$, strengthening the earlier, analytic
arguments, which focused on CHDM with $\Omega_\nu=0.3$.

Quinn, Katz, \& Efstathiou (1996; QKE hereafter) and Thoul \& Weinberg
(1996) find that halos with circular velocities $v_c \simlt
40\;\vunits$ are unlikely to harbor \dla\ absorbers.  Gas in halos below
this limit does not collapse sufficiently to shield itself from the UV
background and reach the necessary HI column densities.  The
shortcoming of the KWHM calculation was that it could not include the
contribution from \dla\ and \lyl\ systems below its resolution limit,
corresponding to a halo circular velocity $v_c \sim 100\;\vunits$.
Consequently, the simulations themselves can only provide a lower
limit to the total amount of \dla\ and \lyl\ absorption in the
Universe.  In Gardner \ea\ (1997a; GKHW hereafter), we addressed this
shortcoming by combining the KWHM results with high resolution
simulations of individual, low mass objects similar to those of QKE.
We used these simulations to obtain a relation between absorption
cross-section $\alpha$ and halo circular velocity $v_c$, which we
combined with the Press-Schechter halo abundance to compute the total
\dla\ and \lyl\ absorption in the SCDM model.  The correction for
previously unresolved halos increased the predicted absorption by
about a factor of two, bringing the predicted \dla\ abundance into
good agreement with observations but leaving the predicted number of
\lyl\ systems substantially below the observed number.  In Gardner \ea\
(1997b; GKWH hereafter), we applied the $\alpha(v_c)$ relation derived
for SCDM to other cosmological models, obtaining more general
predictions for \dla\ and \lyl\ absorption under the assumption that
the relation between halo $v_c$ and gas absorption cross-section was
independent of cosmological parameters.

In this paper, we present results of simulations of several variants
of the inflation+CDM scenario (see, e.g.\ \citealt{khw99}) and improve
upon the GKWH results by using these simulations to predict \dla\ and
\lyl\ absorption in these models.  We continue to use a
Press-Schechter based extrapolation (with the mass function of
\citealt{jenkins01}) to compute the contribution of smaller halos to
\dla\ and \lyl\ statistics, employing an improved methodology that
significantly changes the GKHW predictions for absorption by low mass
systems.  Using an improved fitting procedure, we obtain more accurate
error estimates of our fitted $\alpha(v_c)$ to the simulated data.  We
find that our largest error in estimating the universal amount of \dla\ and \lyl\
absorption arises from the uncertainty in the exact $v_c$ at
which halos cease to harbor these absorbers.  Given the 
large number of halos at $v_{c,min} \approx 40\;\vunits$, a small
variation in the exact value or form of this cutoff leads to
significant deviations in the estimation of total \dla\ and \lyl\ absorption cross
sections.  In light of these results, we find that we are not yet able
to test the four cosmologies we consider against the observed \dla\ 
and \lyl\ abundances.
Instead, we have adopted the approach of 
determining the value of $v_{c,min}$ in each model that yields best
agreement with the observations.

The nature of the galaxies that host \dla\ systems has been a
controversial topic for many years.  Two competing hypotheses have
defined the poles of the debate: the idea that most \dla\ systems are
large, rotating gas disks (e.g.\ \citealt{schiano90}), and the idea that
a large fraction of \dla\ absorption arises in dwarf galaxies (e.g.\
\citet{tyson88}).  The strongest empirical argument for the dwarf
hypothesis is that some imaging studies reveal small galaxies near the
line of sight but no clear candidates for large galaxies producing the
absorption (e.g.\ \citealt{fontana96}; \citealt{lebrun97}; \citealt{moller98}).
The recent study of two \dla\ systems at $z=0.091$ and $z=0.221$ by
Rao \& Turnshek (1998) and Turnshek \ea\ (2000) places especially
stringent upper limits on the luminosities of the host galaxies.  The
strongest argument for the rotating disk hypothesis is the analysis of
metal-line kinematics in \dla\ systems by Prochaska \& Wolfe (1997,
1998), who consider a variety of simplified models for the velocity
structure of the absorbers and find that only a population of cold,
rotating disks with typical circular velocities $v_c \ga 200\;\vunits$
can account for the observed distribution of velocity spreads and for
the high frequency of ``lopsided'' kinematic profiles.  However,
hierarchical models of galaxy formation predict that such massive
disks should be rare at $z \sim 3$.  In an important paper, Haehnelt,
Steinmetz, \& Rauch (1998) showed that hydrodynamic simulations of
high-redshift galaxies could account for the lopsided kinematic
profiles and large velocity spreads found by Prochaska \& Wolfe (1997,
1998) even with halo circular velocities substantially below
$200\;\vunits$, because of large scale asymmetries and departures from
dynamical equilibrium (see also \citet{LP98}).  This result makes the
$\sim 100\;\vunits$ median halo circular velocities found by Kauffmann
(1996) and GKHW for the SCDM model potentially compatible with the
observed metal-line kinematics.  Our present simulations do not yet have
enough resolution for us to repeat the Haehnelt \ea\ (1998) analysis;
we hope to do so with future simulations to carry out a
statistical comparison between results from a randomly chosen
cosmological volume and the Prochaska \& Wolfe (1997, 1998) data.
However, we can already extend the GKHW analysis to other cosmological
models, predicting the fraction of \dla\ absorption arising in halos
of different circular velocities.

We also revisit an important issue explored by KWHM for the SCDM
model, the predicted distribution of projected separations between
\dla\ and \lyl\ systems and high-redshift galaxies.  Given the number
of recent attempts to directly image \dla\ host galaxies, our
predictions will be useful in testing the compatibility of
the size and probable luminosity of our simulated \dla\ hosts with the
imaging data.

Section 2 describes the simulations and our analysis methods.
Section 3 presents our analysis of the \dla\ and \lyl\ systems
resolved by the simulations.  Section 4 describes and applies
our procedures for computing the contribution from unresolved halos.
We discuss the implications of our results and present our conclusions
in \S 5.

\section{Simulations and Methods}
\label{sec:sandm}
\medskip
\subsection{The Simulations}
\label{ssec:Simulation}

\begin{table}
\begin{tabular}{llllllll}
 \tableline\tableline
Name & $\sigma_8$ & $\Omega_m$ & $\Omega_{\Lambda}$ & $h$ &$\Omega_b$ &$n$ &
$M_{res}$
\\ \tableline
\multicolumn{3}{l}{\underline{Principle Runs:}}&&&&& \\
SCDM&   0.7&    1&      0&      0.5& 0.05 &1 & $2.7 \times 10^{11}M_\odot$   \\
CCDM&   1.2&    1&      0&      0.5& 0.05&1 & $2.7 \times 10^{11}M_\odot$   \\
OCDM&  0.75&   0.4&    0&      0.65& 0.03& 1  & $8.2 \times 10^{10}M_\odot$ \\
LCDM&   0.8&   0.4&    0.6&    0.65& 0.03& 0.93  & $8.2 \times 10^{10}M_\odot$\\
TCDM& 0.54&    1&      0&      0.5& 0.05 &0.8 & $2.7 \times 10^{11}M_\odot$ \\
\multicolumn{3}{l}{\underline{Resolution Runs:}}&&&&& \\
L64&    0.8&   0.4&    0.6&    0.65& 0.047& 0.95& $8.2 \times 10^{10}M_\odot$\\
L128&   0.8&   0.4&    0.6&    0.65& 0.047& 0.95& $1.0 \times 10^{10}M_\odot$\\
\tableline\tableline

\end{tabular}
\caption{
\label{tab:params}
Model parameters.}
\end{table}

Our simulations follow the same general prescription as in GKHW, where
a periodic cube whose edges measure 11.11\hinv Mpc in comoving units
is drawn randomly from a CDM universe and evolved to a redshift $z=2$.
We examine the effects of cosmology using five principal simulations
detailed in the first five lines of Table~\ref{tab:params}, where
$\sigma_8$ is the power spectrum normalization, $h \equiv H_0/100$
\kmpersec\ Mpc$^{-1}$, $\Omega_m$ is the fraction of present-day closure
density in matter, $\Omega_b$ is the fraction in baryons, and
$\Omega_\Lambda \equiv \Lambda/(3 H_0^2)$, where $\Lambda$ is the
cosmological constant.  The quantity $n$ is the index of the
inflationary fluctuation spectrum, with $n=1$ corresponding to
scale-invariant fluctuations.  These are the same simulations
presented by Katz et al.\ (1999), who studied the clustering properties
of the galaxies.  We will often refer to the three $\Omega_m=1$ models
(SCDM, CCDM, TCDM) collectively as the ``critical'' models and the two
$\Omega_m=0.4$ models (OCDM, LCDM) as the ``subcritical'' models.

The five principal simulations employ $64^3$ gas and $64^3$ dark matter
particles, with a gravitational softening length of 5\hinv\ comoving
kpc (3\hinv\ comoving kpc equivalent Plummer softening, $1h^{-1}$
physical kpc at $z=2$).  The particle masses are $1.5 \times 10^8
M_\odot$ and $2.8 \times 10^9 M_\odot$ for the gas and dark matter,
respectively, in the critical models and $6.7 \times 10^7 M_\odot$ and
$8.3 \times 10^8 M_\odot$ in the subcritical models.  These
simulations were performed using \tsph\ \citep{hk89}, a code
that unites smoothed particle hydrodynamics (SPH; \citet{lucy77};
\citet{gingold77}) with the hierarchical tree method for computing
gravitational forces \citep{bh86,h87}.

The five principal simulations were done to study the effects of
cosmology on \dla\ and \lyl\ systems.  Because of uncertainties
arising from resolution issues, we add to this study a further two
``next-generation'' simulations, L64 and L128, which were done in the
same cosmology but with different mass resolutions to
investigate the stability of our results with resolution.  These were
performed much more recently using PTreeSPH \citep{SPHINCTR}, a new
parallelized version of \tsph, and their details are also given in
Table~\ref{tab:params}.  L64 is the same mass resolution as OCDM and
LCDM, while L128 is a factor of eight greater in mass resolution
(a factor of two greater
spatially) and is valuable in examining
absorbers in the lower-mass halos that the five principal simulations
cannot resolve.

Detailed descriptions of the simulation code and the radiation physics
can be found in Hernquist \& Katz (1989); Katz, Weinberg, \& Hernquist
(1996; hereafter KWH); and Dav\'e, Dubinski, and Hernquist (1997).  We
only summarize the techniques here.  For both simulation codes, dark
matter, stars, and gas are all represented by particles; collisionless
material is influenced only by gravity, while gas is subject to
gravitational forces, pressure gradients, and shocks.  We include the
effects of radiative cooling, assuming primordial abundances, and
Compton cooling.  Ionization and heat input from a UV radiation
background are incorporated in the simulation.  We adopt the UV
background spectrum of Haardt \& Madau (1996), but reduce it in
intensity by a factor of two at all redshifts so that the mean \lya\
forest flux decrement is close to the observed value given our assumed
baryon density \citep{croft97}.  We apply small further adjustments
to the background intensity during the analysis stage to precisely
match the Press, Rybicki, \& Schneider (1993) measurements of the mean
decrement (see \citealt{croft97} for further discussion of this
procedure).  For example, at $z=3$, the background intensity
is reduced to 20\% of the Haardt \& Madau value,
first by 50\% during the simulation, then by a further 40\% 
during post-processing.  
If we adopted a higher baryon density, $\Omega_b=0.02 h^{-2}$
instead of $\Omega_b=0.0125 h^{-2}$, then the background intensity
matching the observed mean decrement would be a factor $\sim 2.2$ higher.
We use a simple prescription to
turn cold, dense gas into collisionless ``star'' particles.  The
prescription and its computational implementation are described in
detail by KWH.  Details of the numerical parameters can be found in
Katz et al.\ (1999).

\subsection{Halo and Absorber Identification}
\label{ssec:ident}

{}From the simulation outputs at $z=4,$ 3, and 2, we identify dark
matter halos and the individual concentrations of cold, collapsed gas
that they contain.  We initially identify the halos by applying a
friends-of-friends (FOF) algorithm to the combined distribution of
dark matter and SPH particles, with a linking length equal to the mean
interparticle separation on an isodensity contour of an isothermal
sphere with an enclosed average density contrast of $ \delta = 180 $.
Then, the position of the most bound particle in each
FOF-identified halo is passed on to the spherical density method (SO;
\citet{lacey94}) that calculates the sphere about the most bound
particle that contains an overdensity of $\delta=180$.  The halos
used in our analysis are those output from SO, and the circular
velocities, which we denote as $v_c$, are the actual circular
velocities at the $\delta=180$ radius ($R_{180}$) of each halo.
This method of characterizing halo mass has been shown to be the best
for computing the halo mass function \citep{jenkins01}.

To detect discrete regions of collapsed gas capable of producing Lyman
limit and damped \lya absorption, we apply the algorithm of Stadel \ea\
(2001; see also KWH and
http://www-hpcc.astro.washington.edu/tools/SKID) to the distribution
of cold gas and star particles.  SKID identifies gravitationally bound
groups of particles that are associated with a common density maximum.
Gas particles are only considered as potential members of a SKID group
if they have a smoothed overdensity $\rho_g/\bar\rho_g - 1 >
\delta_{vir}$ and temperature $T < 30,000$ K, and we discard groups
with fewer than four members (we will apply a more stringent
resolution cut below).  All of the gas concentrations found by this
method reside within a larger friends of friends halo, even at $z=4$.
We match each absorber with its parent (SO) halo and discard halos
that contain no absorbers.   Including or excluding the ``absorberless''
halos in our mass function does not change the results above our
resolution cutoff (explained below), since nearly all halos above our cutoff
contain at least one absorber.

We calculate the HI column densities for the halos by enclosing each
halo within a sphere centered on the most tightly bound gas particle
and of sufficient size to contain all the gas particles that might
contribute to high column density absorption within the halo.  We
project the gas distribution within this sphere onto a uniform grid
with a cell size of 5.43 comoving kpc, equal to the highest resolution
achieved anywhere in the simulation, using the same spline kernel
interpolation employed by the \tsph\ code for the hydrodynamics.  For
the L128 run, we use a pixel size of 2.715 as the peak spatial
resolution is a factor of two better in each dimension than the other
simulations.  Following KWHM, we calculate an initial HI column
density for each grid point assuming that the gas is optically thin,
then apply a self-shielding correction to yield a true HI column
density (see KWHM for details).  For each halo we compute the
projected area over which it produces damped absorption, with $\NHI >
10^{20.3} \;\cdunits$, and Lyman limit absorption, with $\NHI >
10^{17.2}\;\cdunits$.  For simplicity, we project all halos from a
single direction, although we obtain a similar fit of absorption area
to circular velocity (see below) if we project randomly in the $x$,
$y$, and $z$ directions or average the projections in $x$, $y$, and
$z$.  Projecting a rectangular prism instead of a sphere yields the
same results.  To test for convergence, we reprojected several halos
at 2 and 4 times smaller grid spacings and found that the cross
section for \dla\ and \lyl\ absorption changed by less than 1\% in the
majority of cases and by at most 2.5\%.

\subsection{Numerical Resolution Considerations}
\label{ssec:numres}

Our five principal simulations, which each contain $64^3$ gas and $64^3$
dark matter particles, lack the dynamic range needed to model
simultaneously the full mass range of objects that can contribute to
\dla\ and \lyl\ absorption.  Simulations by QKE and GKHW show that
halos with circular velocities as low as $35\;\vunits$ can host DLA
absorbers, while photoionized gas is unable to collapse and cool in
smaller halos.  However, if we adopted a particle mass low enough to
resolve $35\;\vunits$ halos while retaining the same particle number,
then our simulation volume would be too small to include
a representative sample of more massive halos.

In our analysis of the simulation results, we find that halos
consisting of at least 60 dark matter particles nearly always (more
than 98\% of the time) contain a cold, dense gas concentration.  Below
this threshold, however, a substantial fraction of halos have no cold
gas concentration.  Furthermore, in our bootstrap analysis of the
variance in the relation between halo $v_c$ and absorption
cross-section, described in \S\ref{ssec:resmeth} below, we find much
larger scatter about the mean relation for halos with fewer than 60
dark matter particles than for halos with more than 60 dark matter
particles.

To safeguard against additional systematic effects near the resolution
boundary, we compare the high-resolution L128 run to L64.  We find that
the mass cut of $M_{res}=60(m_{dark}+m_{SPH})$ also allows the L64
halo properties to match smoothly with same mass halos in L128.  Near
perfect agreement for {\dla}s (\cf Figure~\ref{fig:vaL128L64} and later
discussion) is found with a mass corresponding to 70 particles.  So to
be as conservative as possible, we adopt
$M_{res}=70(m_{dark}+m_{SPH})$ as an estimate of the limiting mass
below which we cannot accurately compute the amount of absorption in a
simulated halo.  The 70 particle criterion is more conservative than
the 34 particle criterion that we adopted in GKHW, and this change
will affect our predictions for absorption in lower mass halos in
\S\ref{sec:lowmasshalos} below.

Applying this same mass cut to the L128 run, we find the limiting
resolution to be roughly $50\;\vunits$ at $z=3$, which is
unfortunately still above the mass at which halos cease to host \dla\
and \lyl\ absorption.  Consequently, even the increased dynamic range
of L128 does not allow us to model all absorbers.

In the critical density models, the mass resolution limit is
$M_{res}=2.7\times 10^{11}M_\odot$, corresponding to a circular
velocity at the virial radius of $v_{c,res}=140$, 160, and
$180\;\vunits$ at $z=2$, 3, and 4, respectively.  In the subcritical
models, $M_{res}=8.2 \times 10^{10}M_\odot$, corresponding to
$v_{c,res}=89$, 100, and $112\;\vunits$ at $z=2,$ 3, 4.  In L128, the
mass resolution is a factor of eight better, hence $M_{res}=1.0 \times
10^{10}M_\odot$ and $v_{c,res}=50\;\vunits$ at $z=3$.  For our
statistical analyses of the simulation results below, we always
eliminate absorbers in halos whose total mass (dark matter plus
baryons, with the spherical overdensity mass definition given above)
is $M<M_{res}$.  Our quoted results apply only to halos above the
resolution limit.  In \S\ref{sec:lowmasshalos} we attempt to compute,
as a function of $v_c$, the
contribution of halos with $M<M_{res}$ to the total amount of \dla\
and \lyl\ absorption by combining the Jenkins et al.\ (2001) mass function
with our numerical results.

We assume throughout our subsequent analysis and discussion that our
results for absorption in halos with $M>M_{res}$ are only minimally
influenced by the residual effects of finite numerical resolution.  We
give evidence in \S\ref{sec:lowmasshalos} from the L128 run
that the five principal runs are not influenced by resolution effects
within a factor of 10 of their resolution cutoff.  However, real
absorption systems could have substructure that produces large
fluctuations in HI column density on scales far below resolution limit
of even our highest resolution simulation.  In this scenario, the
total amount of neutral gas in absorption systems would not be very
different from our predictions, except to the extent that clumping
shifts gas above or below the \dla/\lyl\ column density thresholds,
but the distribution of column densities above the thresholds could be
quite different.  This issue will be difficult to address by direct
numerical simulation alone because of the large range of scales
involved.  However, good agreement between predicted and observed
column density distributions would support the contention that the
absorbers do not have a great deal of substructure on scales below the
simulation resolution limits.  KWHM find good agreement between the
predicted and observed shape of the column density distribution in the
\dla\ regime, but a compelling case along these lines will require
simulations that do resolve the full mass range of objects responsible
for damped absorption, and which accurately resolve the low-end cutoff
where halos cease to contain such absorption.

\section{Simulation Results}
\label{sec:simres}

\subsection{Absorption in Collapsed Objects}

\begin{figure}
\plotone{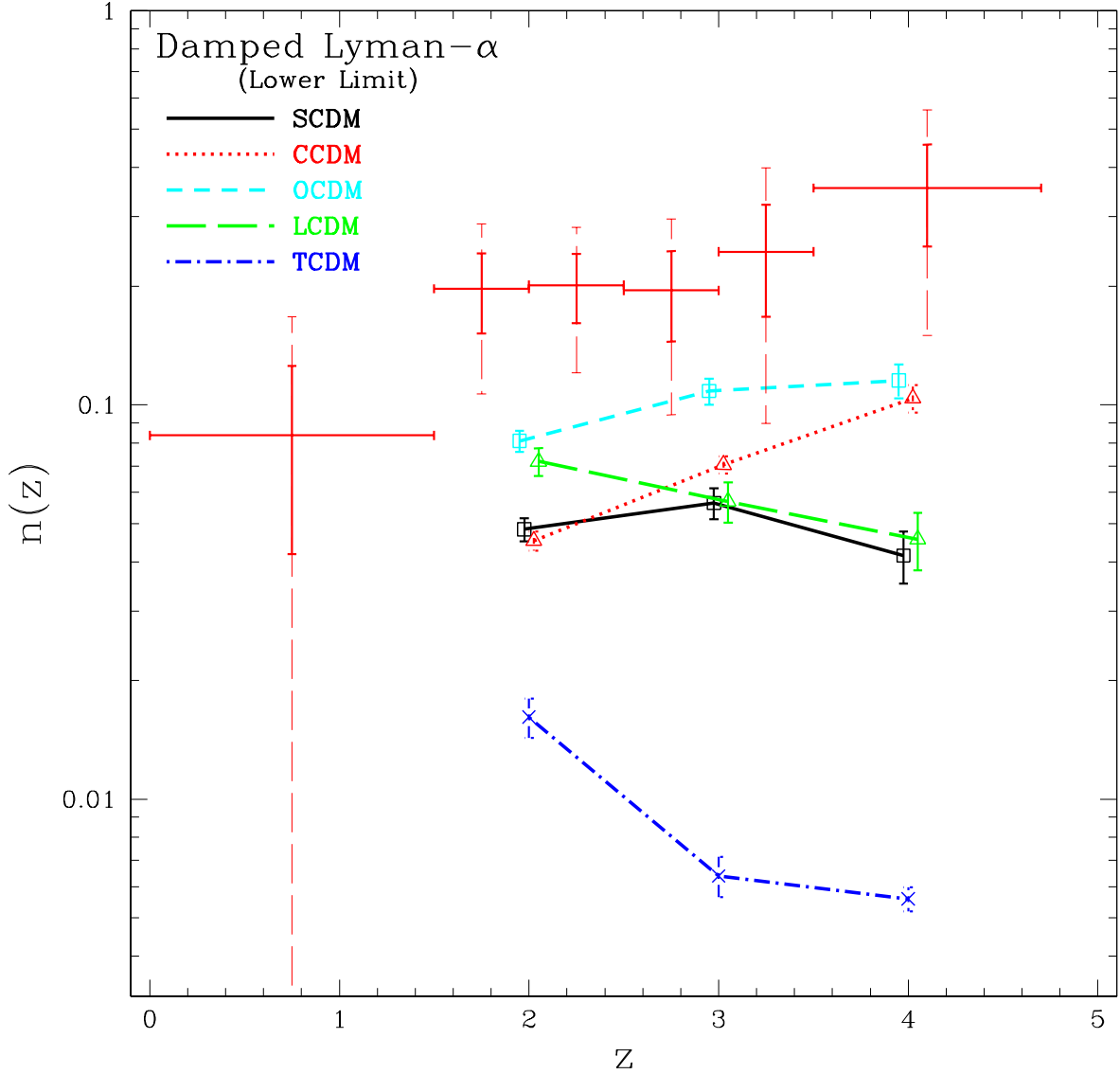}
\caption{
\label{fig:nzloDLA}
The incidence of \dla\ absorption in simulations of the five cosmological
models listed in Table~\ref{tab:params}; $n(z)$ is the number of systems
with $N_{\rm HI}\geq 10^{20.3}\;\cdunits$ intercepted per unit redshift.
Simulation results are computed only for absorption in halos with
mass above the mass resolution limits $M_{res}$ listed in
Table~\ref{tab:params}.
Observational data from Storrie-Lombardi \& Wolfe (2000), shown with
$1\sigma$ (solid) and $2\sigma$ (dashed) error bars, indicate \dla\
absorption by systems of all masses.  The deficiency of absorption
in the models may be explained partly or entirely by the
contribution from halos below the simulations' resolution limits
(see \S\ref{sec:lowmasshalos}).
}
\end{figure}

\begin{figure}
\plotone{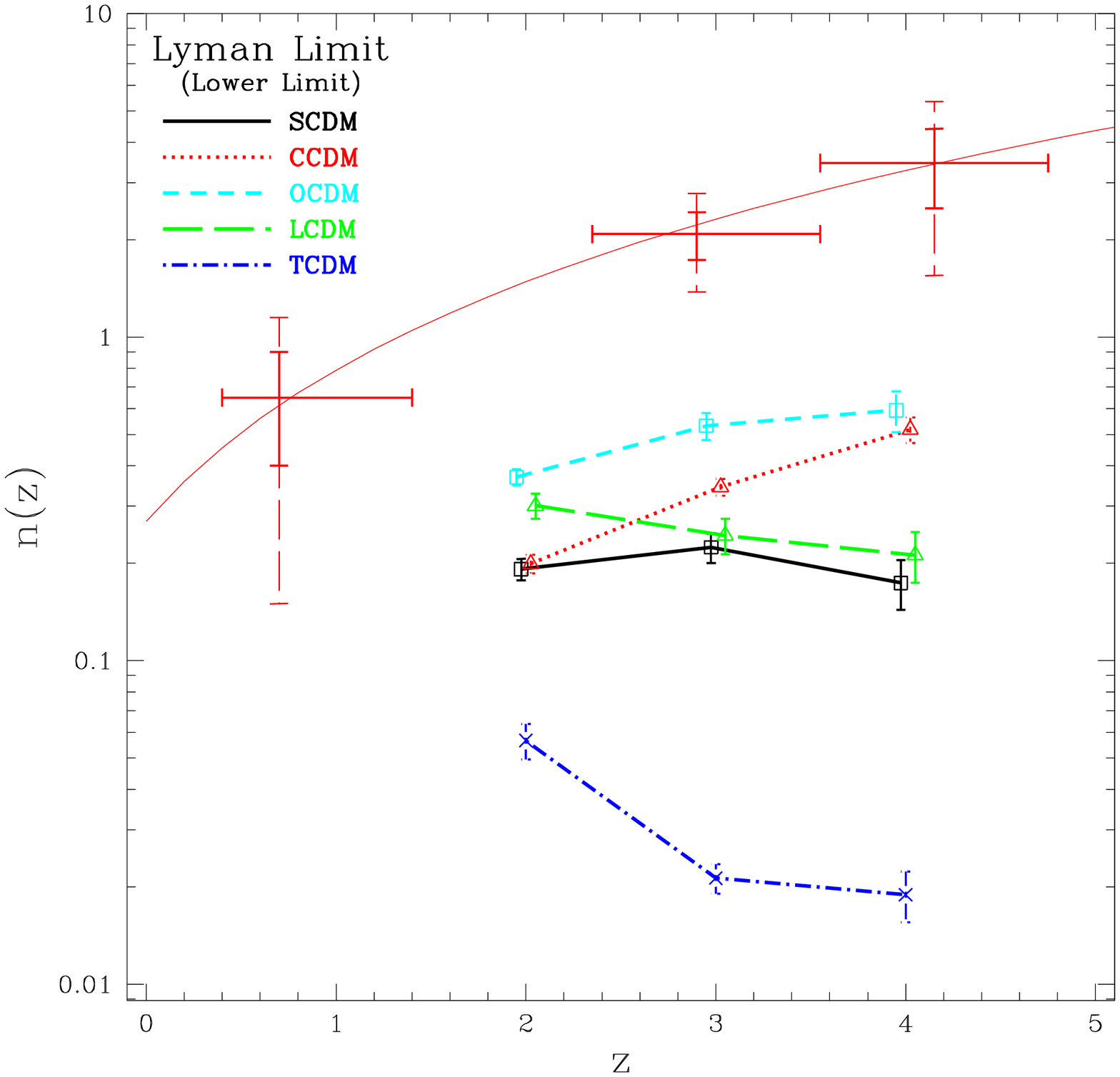}
\caption{
\label{fig:nzloLL}
The incidence $n(z)$ of \lyl\ absorption associated with
halos above the simulation resolution limits.
The format is similar to Fig.~\ref{fig:nzloDLA}, but the threshold
column density is now $N_{\rm HI} \geq 10^{17.2}\;\cdunits$.
The upper error crosses represent the Lyman limit
data of Storrie-Lombardi \ea\ (1994), with $1\sigma$ and $2\sigma$
abundance errors.  The smooth curve shows their fitted power law.
Observational values include systems of all masses.  }
\end{figure}

Figures \ref{fig:nzloDLA} and \ref{fig:nzloLL} show the incidence of
\dla\ and \lyl\ absorption in our five cosmological models: $n(z)$ is
the mean number of absorbers intercepted per unit redshift above the
\dla\ (Fig.~\ref{fig:nzloDLA}) or \lyl\ (Fig.~\ref{fig:nzloLL}) column
density threshold.  The numerical results for halos above the mass
resolution limit $M_{res}$ are shown at $z=2,$ 3, and 4.
Observational results for \dla\ absorption are taken from
Storrie-Lombardi \& Wolfe (2000; \cf also Storrie-Lombardi, Irwin, \&
McMahon 1996a; Wolfe \ea\ 1995) and for \lyl\ absorption from
Storrie-Lombardi \ea\ (1994).  When comparing the subcritical models to
the critical models, note that the subcritical models have lower
$M_{res}$ and therefore sample the distribution of absorbers down to a
lower mass cutoff, boosting the $n(z)$ prediction relative to that of
the critical models.  Taken directly from the simulations and from
halos only above $M_{res}$, the values in these plots are hard lower
limits to the predicted $n(z)$.  The limiting circular velocities
$v_{c,res}$ are below the value $v_c \sim 120\;\vunits$ inferred
by Prochaska \& Wolfe (1997, 1998) for typical \dla\ circular
velocities based on a rotating disk model for metal-line kinematics,
and even so the predicted number of \dla\ absorbers is usually a
factor of two or more below the observed value.  We conclude that if
the inflationary CDM models considered here are even approximately
correct, then the asymmetries and large velocity spreads found by
Prochaska \& Wolfe must be a result of complex geometry and
non-equilibrium dynamics, as proposed by Haehnelt et al.\  (1998).

\begin{figure}
\plotone{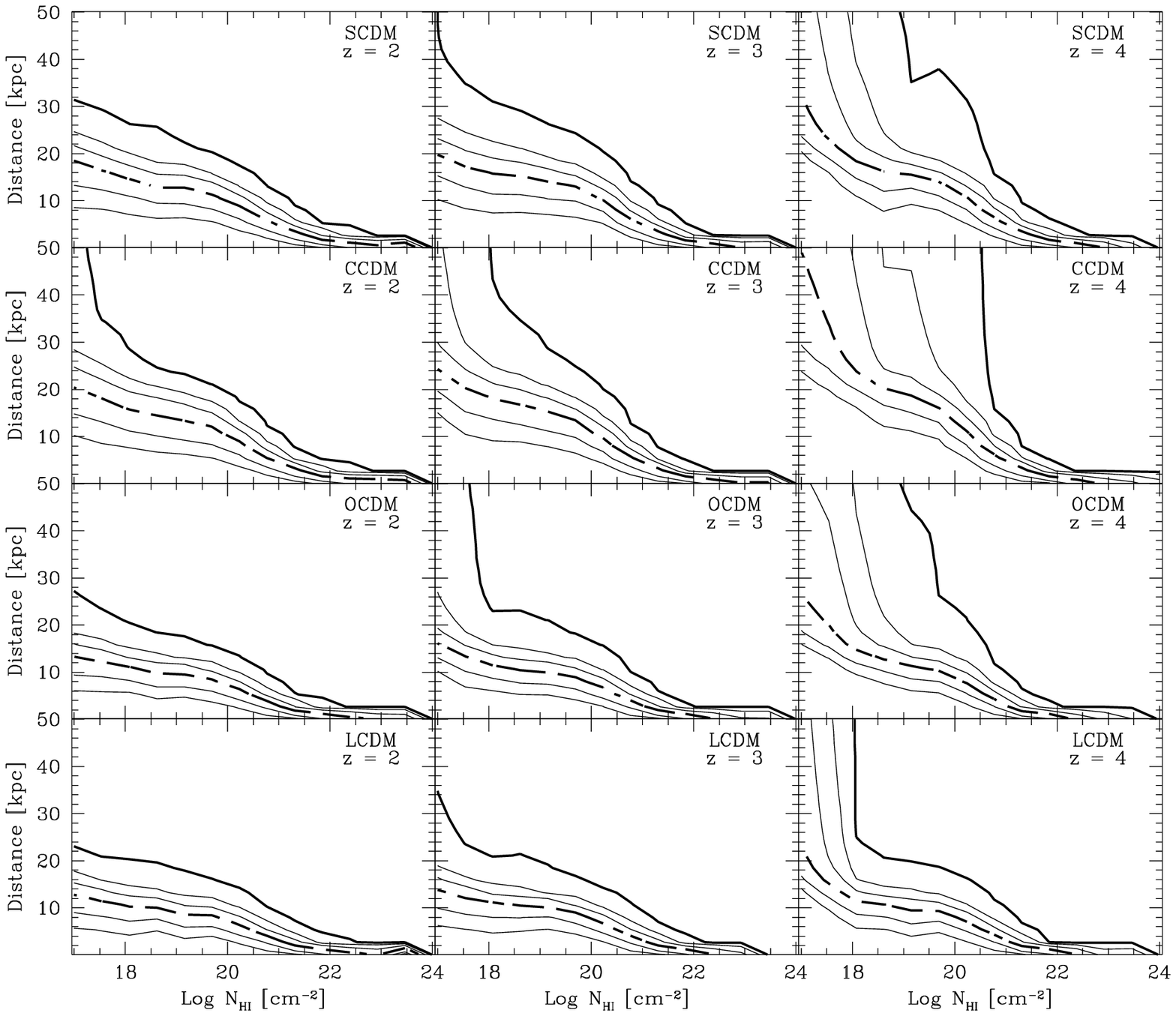}
\vglue-1.0in
\caption{
\label{fig:impacts}
The distribution of impact parameters $D_{\rm proj}$ from high
column density lines of sight to neighboring galaxies.
The $x$-axis denotes the HI column density along the
line of sight.  The $y$-axis is the projected distance $D_{\rm proj}$
(in physical kpc) to the galaxy
in the simulation volume that lies closest (in projection)
to the line of sight.
The contours indicate the percentage of lines of sight containing an
absorber of HI column density $\NHI$ that have a galaxy
whose center is within
$D_{\rm proj}$.  The thin contour levels
are 10\%, 25\%, 75\% and 90\% with the dashed thick line
denoting 50\% and the solid thick line 99\%.
To construct the contours of the plot,
lines of sight in the simulations were sorted into bins
$\Delta\NHI=0.5$ dex and $\Delta D_{\rm proj}=2.5$ kpc wide.}
\end{figure}

\begin{figure}
\plotone{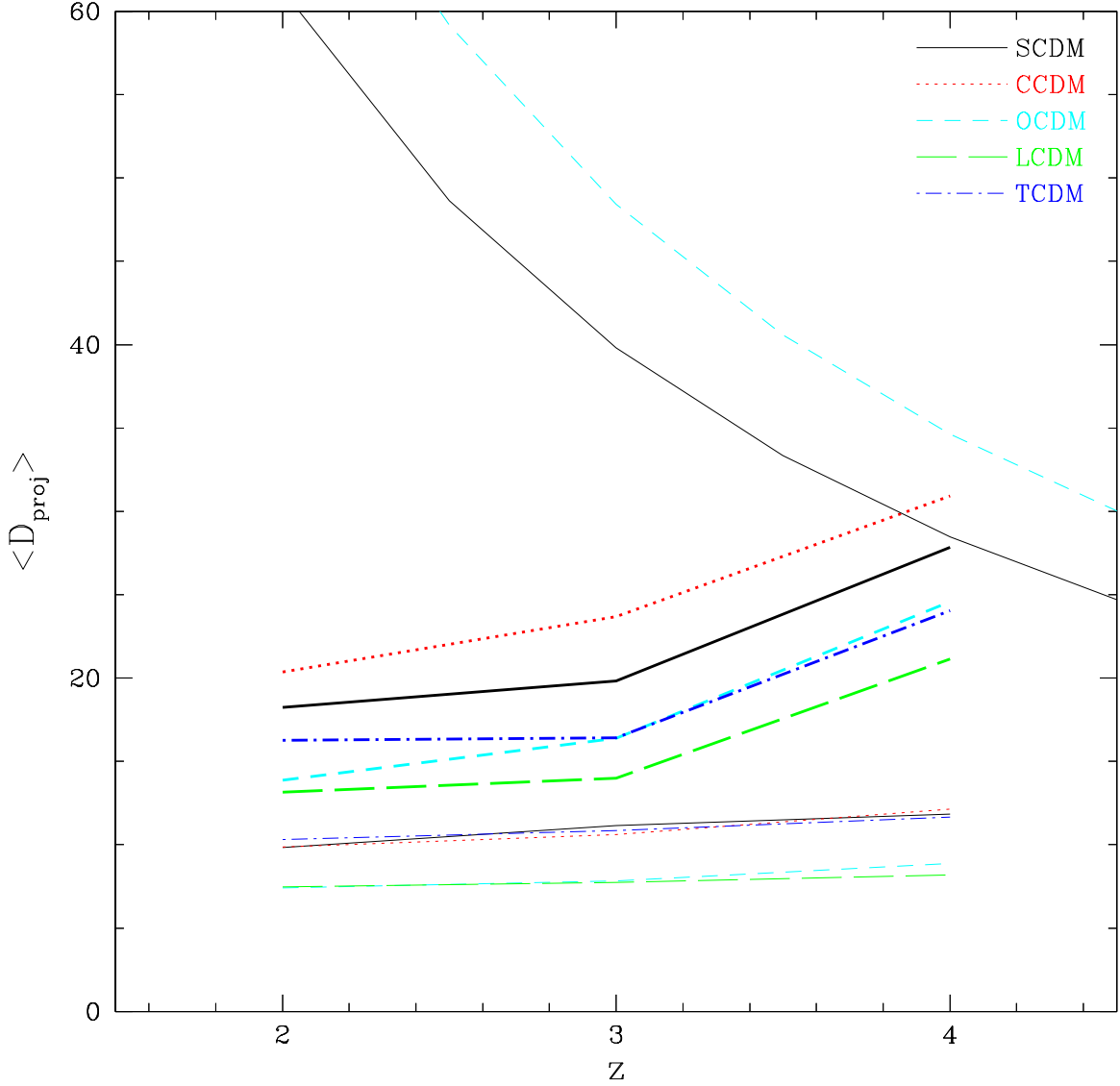}
\caption{
\label{fig:impactmeans}
Average impact parameter (in physical kpc) vs.\ redshift.  The bold
lines are the average impact parameter $\langle D_{proj} \rangle$ 
of lines of sight
with $10^{17} \leq \NHI \leq 10^{17.5}\;\cdunits$.  The lower thin set of
lines is the same for $10^{20} \leq \NHI \leq 10^{20.5}\;\cdunits$.  The two
curves at the top of the plot show the virial radius of a 150 km/s
halo in a $\Omega_m = 1$ universe (solid) and $\Omega_m = 0.4$ universe
(dashed).}
\end{figure}

Figure~\ref{fig:impacts} shows the distribution of impact parameters
$D_{\rm proj}$, in physical units, between high column density
absorbers and the centers of neighboring galaxies.  Specifically, the
contour level represents the percentage of lines of sight of a given
$\NHI$ for which the closest simulated galaxy, in projection, lies
within a projected distance $D_{\rm proj}$.  Note that in this figure
and in subsequent figures we represent distance in kpc and not \hinv\
kpc.  Nearly all of the high column density systems in our simulations
are associated with a galaxy, with the highest column density systems
sampling the innermost regions of the galaxy and the lower column
density systems occurring at larger impact parameters.  At $z=2$,
nearly all \dla\ systems lie within 15-20 kpc of a galaxy center, and
nearly all \lyl\ systems lie within 30 kpc.  At higher redshifts, the
most likely impact parameter increases, which could indicate a
physical contraction of \dla\ systems as they age or could
alternatively reflect the higher neutral fraction associated with a
given overdensity at higher redshift (similar to the interpretation of
evolution of the low column density forest given by \citet{HK96} and
\citet{Dave99}).  This increase can easily be seen in
Figure~\ref{fig:impactmeans}, which plots the mean impact parameter for
systems at the \dla\ and \lyl\ cutoff and compares them with the
virial radius of a $v_c = 150\;\vunits$ halo.  The mass, and therefore
size, of an isothermal sphere with a given circular velocity goes as
$(1+z)^{-1.5}$.  Not only does the mean absorption cross section
increase at redshift $z > 3$, but the size of the parent halos
decreases, meaning that at $z=4$ the fraction of the area of the halo
subtended by \dla\ and especially \lyl\ absorption is much larger than at
$z=2$.  We will further examine absorber area vs.\ halo virial radius
in Section \ref{ssec:resmeth}.

\begin{figure}
\plotone{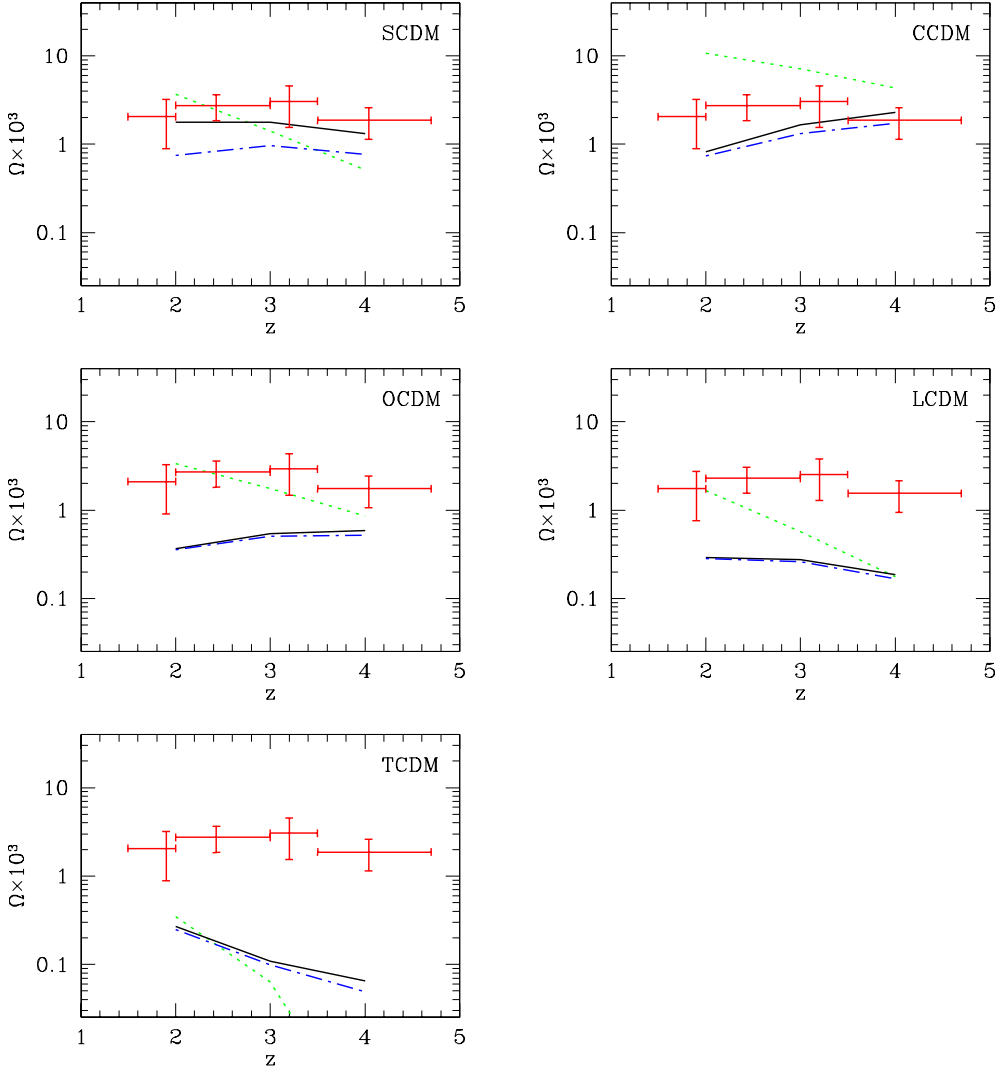}
\caption{
\label{fig:omegaccg}
Fraction of critical density in cold collapsed gas and stars in
halos with $v_c \geq v_{c,res}$.  The solid lines are the density
parameter $\Omega_{ccg}$ in cold collapsed gas
for each cosmological model.  The dotted lines
are $\Omega_*$, the same quantity for the stars.  The
dot-dashed lines show ``observational'' values of $\Omega_{ccg}$,
which include
only lines of sight with HI columns $\NHI < 10^{21.8}$ cm$^{-2}$.  The
$1\sigma$ error crosses are taken from Storrie-Lombardi
et al.\ (1996b) and adjusted for the appropriate cosmology.}
\end{figure}

Figure~\ref{fig:omegaccg} shows the fraction of critical density in
cold collapsed gas, $\Omega_{ccg}$ (solid line), and the fraction of
the critical density in stars, $\Omega_\star$ (dotted line), as a
function of redshift in the various cosmological models.  In
subcritical models, we define $\Omega_x(z) \equiv \rho_x(z) \times
(1+z)^{-3}/\rho_c(z=0)$, i.e., $\Omega_x$ represents the comoving
density of component $x$ relative to the critical density at $z=0$.
We obtain $\Omega_{ccg}$ by integrating the column density
distribution $f(\NHI)$ for all of the halos in the simulation.  Error
crosses show the values derived by Storrie-Lombardi, McMahon, \& Irwin
(1996b) from a sample of \dla\ systems.  The largest observed column
density for \dla\ systems in the Storrie-Lombardi et al.\ (1996ab)
sample is $10^{21.8}$ cm$^{-2}$, probably because higher column
density systems are too rare to have been detected. For a more direct
comparison to the data, we therefore compute an ``observational''
value, $\Omega_{obs}$, for which we only count gas along lines of
sight with $\NHI \leq 10^{21.8}\;\cdunits$.  The contribution to
$\Omega_{ccg}$ from higher column density systems is generally small,
but it is significant in the SCDM model.  In all cases, we include
only gas in halos with $v_c \geq v_{c,res}$.

For all the models in Figure~\ref{fig:omegaccg}, $\Omega_\star$ increases
steadily as the Universe evolves.  However, $\Omega_{ccg}$
remains constant to within a factor of two from redshift
4 down to redshift 2, indicating that additional gas cools and
collapses to replace the gas that is turned into stars.  Gas reaches
higher densities in the the SCDM and CCDM models, leading to a
larger difference between $\Omega_{ccg}$ and $\Omega_{obs}$.
In nearly all cases, our ``observed'' cold gas densities $\Omega_{obs}$
fall at least a factor of two below the observational data
of Storrie-Lombardi et al.\ (1996b), which could themselves be
underestimates of the true cosmological values of $\Omega_{ccg}$ if dust
extinction is important \citep{pei95}.
If any of these models are to be viable, a substantial fraction of
the high redshift HI must reside in systems below our resolution limit,
an issue to which we turn in \S\ref{sec:lowmasshalos}.

To obtain the simulation values of $\Omega_{ccg}$ in Figure~\ref{fig:omegaccg},
we had to alter the procedure described in \S\ref{ssec:ident} for
computing HI column densities.
While our standard grid spacing of 5.43 comoving kpc is sufficient to resolve
objects with HI columns of $10^{20.3}$ cm$^{-2}$ and lower, a finer
mesh is required to resolve the cold dense knots of gas at higher
column densities, which contribute significantly to the $\Omega_{ccg}$
integral.  As described in KWHM, we generate the initial HI map assuming
complete transparency, then use the mass, HI mass fraction,
and temperature of each grid cell to correct the HI column density for its
ability to shield itself from the surrounding radiation.
Typically, a high column density grid cell contains some
regions where the hydrogen should be partly ionized and some where
it should be completely neutral owing to self-shielding effects.
At the standard
resolution of 5.43 comoving kpc, our procedure may average the
contributions of these two regions before the self-shielding
correction is applied, resulting in a lower neutral column density than if the
identical correction procedure were applied with a smaller grid
spacing.  This effect is not important for computing $n(z)$, the
number of systems with $\NHI$ above the \dla\ cutoff, but it
can be important for computing the {\it total} mass density of neutral
gas, $\Omega_{ccg}$.  We examined the
effect by reprojecting some of the halos at z=2,3,4 in the SCDM model
at 2 and 4 times the original spatial resolution.  The original
resolution of 5.43 kpc underpredicts the total HI in the simulation,
while the cold collapsed gas mass in the 2X and 4X cases is nearly
identical.  Consequently, we regard the 2X case as numerically converged.

Unfortunately, reprojecting all the simulation outputs at this higher
resolution is not computationally feasible.  We therefore developed an
approximate procedure based on the original grid spacing, calibrated
against the few higher resolution SCDM projections.  In grid cells
where the self-shielding corrected HI column is greater than a
threshold value $N_{{\rm HI},c}$, we treat as fully neutral
all gas particles that contribute to
that grid cell and meet the following criteria:
temperature $T<30,000$K and gas density
$\rho_g > (1000/177) \rho_{vir} (\Omega_b/\Omega_m)$,
where $\rho_{vir}$ is the virialization overdensity
described in \S\ref{ssec:ident}.
For critical models, the density cut corresponds to $1000\ \Omega_b$.  In
subcritical models, the density cut occurs at the same fraction of the
critical density as in the $\Omega_m=1$ models.
We find that for $\log N_{{\rm
HI},c} = (20.4, 20.7, 20.7)$ at $z=(2,3,4)$ this procedure reproduces
the SCDM high resolution values for $\Omega_{ccg}$ and $\Omega_{obs}$
to within 10\%.

\subsection{Other Possible Sources of Absorption}

It is possible that some Lyman limit and/or \dla\ absorption
originates from regions other than galactic halos.  To investigate
this alternative within our simulations, we project the entire
simulation volume and compare the area of \lyl\ and \dla\ absorption
to the sum of the absorption calculated by projecting each halo
individually.  In the analysis presented here, we
use all halos that have at least one group identified by SKID as
described in \S\ref{ssec:ident} (\ie at least one concentration
of cold gas that is gravitationally bound), whether or not the halo
itself has $M\geq M_{res}$.
Above $M=M_{res}$, 98\% of the dark matter halos harbor at least one
SKID-identified group.

We have removed TCDM from the analysis in this section due to the extreme
paucity of structure in the model.
For the remaining four models, we calculate the total area subtended by \dla\
absorption in the halos with SKID-identified groups.  Comparing this
value to the total area subtended in an entire volume projection of
each simulation at redshifts $z=2$ and $z=4$, we find agreement within
4.5\% for all the models at both redshifts, and to better than 2\% in
five of the eight cases.  We attribute the remaining differences to
having more than one absorber along a given line of sight.  Hence, all
\dla\ absorption in the simulation occurs within halos with at least
one concentration of cold, gravitationally-bound gas.

In \lyl\ absorption, five of the eight outputs agree to
better than 6\% when compared in this manner.  However, at
$z=4$ the results of volume projection and halo projection
differed by 15\%, 30\%, and 13\% for SCDM, CCDM, and OCDM
respectively.  We took
the worst case, $z=4$ CCDM, and projected all the halos that contained at
least 32 particles (gas $+$ dark matter), whether or not they contained a
SKID-identified gas concentration.
When we sum the area subtended by
\lyl\ absorption in these halos, we find that it now
accounts for all but 1.2\% of the \lyl\ absorption found by
projecting the entire volume.  Hence Lyman limit absorption still
occurs exclusively in halos, but in this instance 30\% of it occurs
in halos in which our resolution of gas dynamics and cooling is only marginal
and in which there are no SKID-identified gas concentrations.
If our study had higher resolution,
it is likely that some of these
halos
would have been able to
form \dla\ systems as well.  However, it is just to correct for these unresolved
or under-resolved halos that we developed our Press-Schechter correction
technique. The important conclusion is that all the Lyman
limit absorption we find in the simulations resides within dark matter
halos.  If Lyman limit absorption were to occur outside galactic halos, it
would have to be in regions that are much too small for us to resolve.

To summarize, all \dla\ and \lyl\ absorption in our simulations
occurs in dynamically bound dark matter halos, even below our
resolution cutoff.  At $z=2$ and $z=4$, in all four of the models
tested, \dla\ absorption arises entirely in objects identified by SKID
as bound concentrations of cold gas.  \lyl\ absorption in the simulations
also occurs exclusively in dark matter halos, although
at $z=4$ some of the gas within these halos
is
able to reach \lyl\ column densities without becoming
a SKID-identified concentration.

\section{\dla\ and \lyl\ Absorption by Low Mass Halos}
\label{sec:lowmasshalos}

\subsection{Motivation from Higher Resolution Simulation}
\label{ssec:resmeth}

\begin{figure}
\plottwo{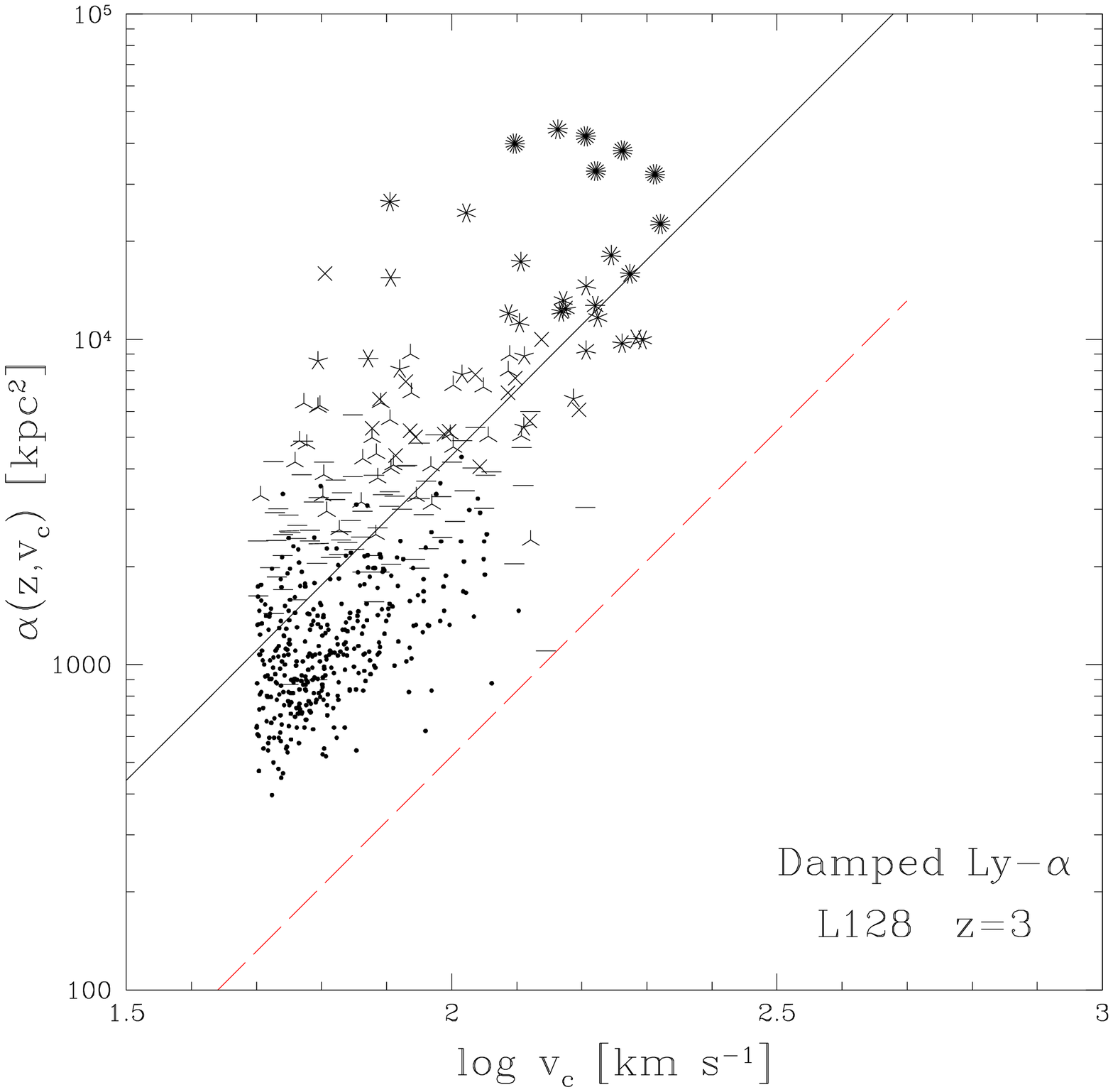}{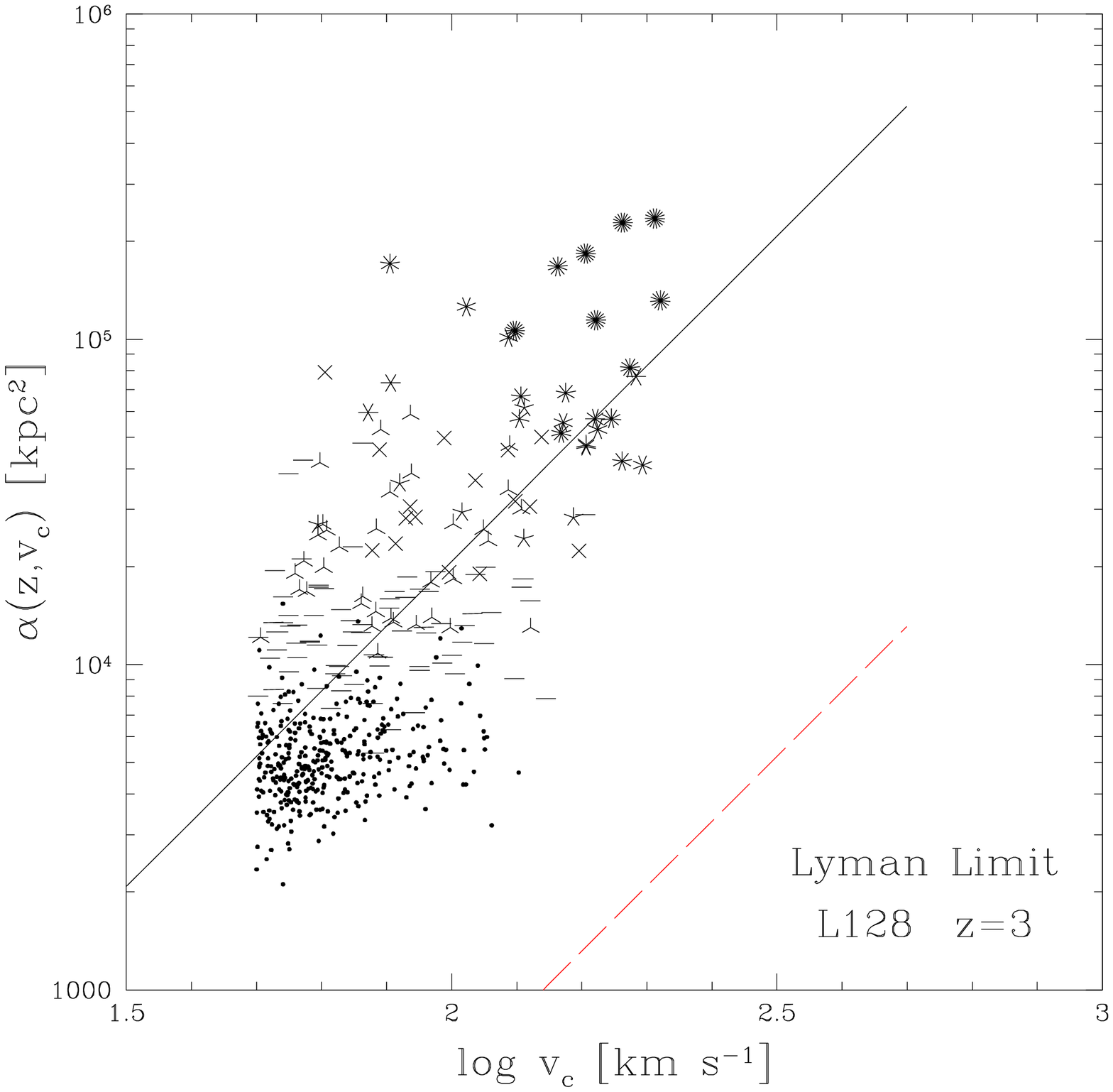}
\caption{
\label{fig:vaL128}
Comoving absorbing area (kpc$^2$) vs. circular velocity $v_c$
(\kmpersec) for halos in the L128 run at $z=3$.  The left and right
panels show the area subtended by \dla\ and \lyl\ absorption
respectively.  The number of vertices in each data point corresponds
to the number of gas concentrations in the halo, with the solid points
representing halos containing a single absorber.  In the left (\dla)
panel, the solid line is the area subtended by a face-on disk of
radius $R = 0.29 R_{vir}$, where $R_{vir}$ is the virial radius of an
isothermal sphere with circular velocity $v_c$ at the virial radius.
In the right (\lyl) panel, the solid line corresponds to a a face-on
disk of radius $R=0.63 R_{vir}$.  The dashed line is a face-on disk of
radius $R = 0.1 R_{vir}$.  }
\end{figure}

We have so far focused on \dla\ and \lyl\ absorption in halos above
our simulation mass resolution limits $M_{res}$.  However, if we want
to test cosmological models against the observed incidence $n(z)$, we
must also consider the absorption that arises in lower mass halos,
which are smaller in cross section but much more numerous.  The L128
run is a factor of eight finer in mass resolution than the other
simulations in this study, allowing us to examine trends in \dla\ and
\lyl\ systems to lower masses.  Figure~\ref{fig:vaL128} compares, at $z=3$, the
circular velocity at $R_{180}$ of the halos in L128 with the cross
section subtended by \dla\ absorption (left panel) and \lyl\
absorption (right panel) when the halo is projected.  The number of
vertices on each point indicates the number of SKID-identified
concentrations of cold collapsed gas within the halo.  We can see that
the absorption characteristics of galactic halos are not well
approximated by assuming a single galaxy per halo.  Instead, the correlation of
absorption cross section $\alpha(z,v_c)$ with halo mass seems to arise
not from a single galaxy in each halo becoming larger as its parent
halos increases in mass, but rather from more massive halos harboring
more galaxies.  Consequently, the multiple-absorber nature of halos is
extremely important in modeling the connection between halo mass and
absorption cross section.  To approach the problem semi-analytically,
it is necessary to model the full interaction history of the halos, as
is done in Maller et al.\ (2000).  The
absorption cross sections of the individual galaxies is virtually
independent of halo mass.  If any trend exists, it appears that
$\alpha(z,v_c)$ in halos containing only one galaxy may actually decrease
slightly in more massive halos.
Higher mass halos have deeper potential wells, causing the
concentrations of cold gas to contract more efficiently.  This complex
gas dynamical behavior demonstrates the value of a fully numerical
treatment in modeling these objects.

The dashed lines in Figure~\ref{fig:vaL128} show the area subtended by
a face-on disk of radius $R=0.1 R_{vir}$ where $R_{vir}$ is the virial
radius of an isothermal sphere with circular velocity $v_c$ at the
virial radius.  10\% of the virial radius is the typical extent of a
galaxy based on centrifugal arguments.  The solid line in
Figure~\ref{fig:vaL128} left (\dla) panel shows the area subtended by
a face-on disk of radius $29\% R_{vir}$, while in the right panel it
denotes area $\alpha=\pi (63\% R_{vir})^2$.  Although the solid lines were not
fit to the data, one can see that the general trend of \dla\ and \lyl\
absorption is that the cross-section is roughly described as
proportional to $(30\% R_{vir})^2$ and $(60\% R_{vir})^2$ respectively.

\subsection{Testing Extrapolations to Lower Mass Halos}

We would like to be able to extrapolate the contribution to the total
incidence by halos with $v_{c,min} < v_c < v_{c,res}$.  Given the
results from Section \ref{ssec:resmeth}, it is plausible to assume a
power law fitting function $\alpha_{\rm PL}(z,v_c) \equiv A v_c^B$.
QKE find that a photoionizing background suppresses the collapse and
cooling of gas in halos with circular velocities $v_c < v_{c,min} =
37$ \kmpersec.  Thoul \& Weinberg (1996) find a similar cutoff in 
simulations that are much higher resolution but assume spherical symmetry.
As discussed
in \S\ref{ssec:Simulation}, $v_{c,res}$ is approximately
$140\;$\kmpersec\ in the critical models and $89\;$\kmpersec\ in the
subcritical models at $z=2$ for the five principal runs.  In L128, which
has a resolution limit of $v_{c,res} \approx 50\;$\kmpersec, we find
no evidence of a photoionization cutoff; simulations with better mass
resolution are required to detect it.  
For the $\alpha(v_c)$ dependence we find for resolved halos in our
simulations, low mass halos dominate the total cross-section for \dla\ 
and \lyl\ absorption.  Therefore, the predicted incidence $n(z)$ depends
sensitively on the assumed value of $v_{c,min}$.

Since we cannot robustly predict $n(z)$ without exact knowledge of 
$v_{c,min}$, we adopt the less ambitious goal of determining, for each
cosmological model, what value of $v_{c,min}$ yields a good match
to the observed values of $n(z)$.  
Our approach to this calculation is similar to that of GKHW:
we use our numerical
simulations to calibrate a fit to the mean cross section for \dla\ (or
\lyl) absorption of halos with circular velocity $v_c$ at redshift
$z$, then integrate over an analytic halo mass function to compute $n(z)$.

Figure~\ref{fig:vaL128L64} compares $\alpha(v_c)$
in the L128 and L64 simulations at
redshift $z=3$.  Figure~\ref{fig:nzplotsim} compares the cumulative
incidence $n(z,v_c)$, the total incidence from absorbers in
halos at least at massive as $v_c$, in L128 and L64.  Note that in
this Figure, the incidence is measured directly from the simulations
and not by convolving an analytic mass function with $\alpha_{\rm
PL}(z,v_c)$ as is done later in this section.  The runs were performed
using the same cosmology but a factor of eight different mass resolution.
In the absence of systematic resolution effects that may exist above
our cutoff $v_{c,res}$, the data from the L64 simulations should
smoothly overlap with L128 above the L64 $v_{c,res}$ value.  For the
\dla\ case, although L64 appears to have more low-mass outliers than
L128, the simulations agree quite well both in $\alpha(v_c)$ and
incidence.  In the \lyl\ case, however, we find that the we
systematically underestimate the absorption cross section of \lyl\
absorbers in L64 by roughly 25\%.  This leads to the cumulative
incidence of the L64 simulation also being 25\% less than the same
$v_c$ in L128.  Therefore, at the L64 resolution we are not resolving
all \lyl\ absorption regions that exist inside halos with $v_c \geq
v_{c,res}$.  One possibility is that areas whose average HI column
density is slightly below the \lyl\ cutoff ($\NHI \geq
10^{17.2}\;\cdunits$) on $\sim 5$ kpc scales have smaller, patchy
clumps with higher column density.  Hence, the lower resolution
simulations would systematically underestimate the \lyl\ absoption in
these regions.  For \dla\ absorption, however, we detect no signatures
of numerical resolution artifacts.

To find the best fitting function for absorption cross section, we bin
halos in 0.05 dex increments in $\log v_c$, beginning with
$v_{c,{res}}$ and subject to the constraint that there be at least 10
halos in each bin.  We sometimes are forced to widen the bin size to
satisfy the latter constraint.  Let $\sigma_{DLA}=\alpha(v_c,z)$ 
denote the ``cross
section'' of \dla\ absorption, i.e. the comoving area subtended by HI
column densities $\NHI \geq 10^{20.3} \;\cdunits$ when a halo is
projected onto a plane.  For the binned distribution of halos, we
determine the log of the average halo \dla\ absorption cross section,
$\log \langle\sigma_{DLA}\rangle$ for each bin.  Then we calculate the
statistical uncertainty of $\log \langle\sigma_{DLA}\rangle$ in each
bin by using the bootstrap method with 1000 random realizations of the
data set of halos with $M \ge M_{res}$.  The distribution of halo
cross sections $\sigma_{DLA}$ at a given $v_c$ is approximately
log-normal and hence best described by a Gaussian in log space.  Since
we will use the bootstrap errors in the next section to calculate
confidence limits, which assume Gaussianity, we express the errors in
log space.  We fit the points $\log\langle\sigma_{DLA}\rangle$ by
linear least squares to determine the parameters $A$ and $B$, the
amplitude and index for the power law fitting function $\alpha_{\rm
PL}(z,v_c) \equiv A v_c^B$.  The error crosses in
Figure~\ref{fig:vaL128L64} show the mean absorption cross section in
each bin of circular velocity at redshifts $z=3$ for \dla\ and \lyl\
systems.  The horizontal error bars show the width of each bin, and
the vertical error bars show the $1\sigma$ logarithmic uncertainty in
$\langle\sigma_{DLA}\rangle$ for the bin determined by the bootstrap
procedure.  The solid error crosses are for L128 data and dashed
error crosses for L64.  The solid and dashed lines denotes the best fit to
the L128 and L64 data respectively.  For \dla\ systems, the fit to the
L64 data is nearly identical to the L128 fit, showing that data at the
resolution of the principal runs can be used reliably to estimate
the absorption cross section below their resolution cutoff.  The \lyl\
fit is systematically lower, reflecting the mismatch between L128 and
L64 halo absorption.  However, the two lines parallel each other
meaning that the fitting procedure is robust and allows an accurate
extrapolation to lower halo masses.

It is important to note that we are not seeking to model the
{\it distribution} of absorption cross sections in each bin,
but only to characterize the {\it mean} cross section in each bin
so that the total absorption can be accurately reconstructed
from equation~(\ref{eq:nofzM}) below. The large spread and asymmetry in
the distribution of absorption cross sections for individual halos
is inconsequential for our purposes: $\alpha_{\rm PL}(v_c,z)$ is the number
which, when multiplied by the halo number density at $v_c$, yields the
total absorption at that $v_c$ that matches the total absorption
present in the simulation.  The bootstrap technique yields a robust
estimate of the statistical uncertainty in this mean cross section
caused by the finite number of halos in the simulation.
Although the spread in absorption cross sections of individual halos
may be large, their average absorption is a well determined quantity.

\subsection{Results}

\begin{figure}
\vglue-0.65in
\plotone{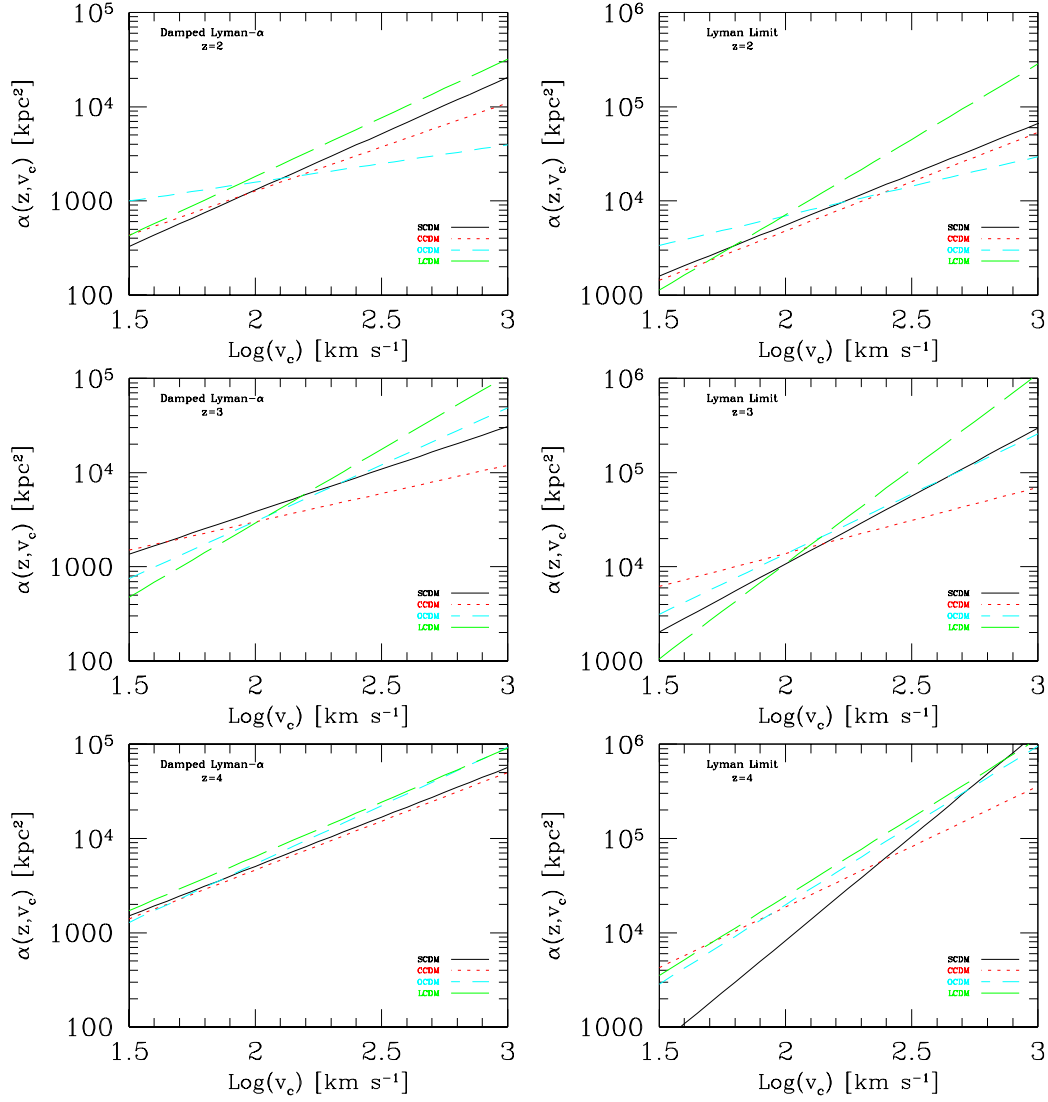}
\vglue-0.26in
\caption{
\label{fig:fitplots}
The best-fit absorption cross sections $\alpha(v_c,z)$ for \dla\ and
\lyl\ systems in each model.  The area is given in comoving kpc$^2$. }
\end{figure}

Figure~\ref{fig:fitplots} shows the fitted relation $\alpha_{\rm
PL}(z,v_c)$ for each of 4 cosmologies and redshifts $z=(2,3,4)$.  Again we
have removed TCDM from the analysis in this section due to the extreme
paucity of structure in the model.  Given so few halos above the mass
limit, we felt it useless to attempt to fit to the data.
The values of the fit parameters are detailed in Table~\ref{tab:fits}.
At $z=4$, the fits for each model tend to follow the same slope (the
exception being \lyl\ SCDM).  At later redshifts, $\alpha_{\rm
PL}(z,v_c)$ evolves differently for different models.  
In general, LCDM tends to be among
the steepest in all cosmologies.  Interestingly, this steepness is not
paralleled by OCDM at $z=2$ where the cross-sections of the more
massive halos in OCDM have decreased markedly.  CCDM tends to be
flatter than SCDM --- the increased amplitude of structure apparently results in
halos having smaller gas cross-sections.  It is difficult to draw any general 
conclusions about the behavior of high column density absorbers in
critical models vs. subcritical models, except that LCDM tends
to have the highest cross-sections in massive halos.

We are now in a position to correct the $n(z)$ estimates from
\S\ref{sec:simres} to include the contribution from halos with
$M<M_{res}$.  Our approach to this problem is similar to that of GKHW,
though there are important differences of detail that make a
significant difference to the end results, as we discuss later.  We
compute the number density of halos $N(M,z)$ as a function of mass at
specified redshift using the mass function of Jenkins \ea\ (2001; see their
equation 9).  
Multiplying $N(M,z)$ by our numerically calibrated functions
$\alpha_{\rm PL}(v_c,z)$, and integrating from $v_c$ to
infinity, yields the number of \dla\ (\lyl) absorbers per unit redshift
residing in halos of circular velocity greater than $v_c$:
\be
n(z,v_c)={dr \over dz} \int_{M(v_c)}^{\infty} N(M',z) \alpha_{\rm PL}(v_c',z)
dM' ,
\label{eq:nofzM}
\ee
where $v_c'$ is the circular velocity at $R_{180}$, the
$\delta=180$ radius, of a halo of mass $M'$, and $r$ is
comoving distance (see GKHW for detailed discussion).  If one takes
the lower limit of the integral to be $v_{c,min}$, the minimum
circular velocity for gas cooling and condensation, this yields
$n(z)$, the total incidence of \dla\ (or \lyl) absorption at redshift
$z$.

\begin{figure}
\plottwo{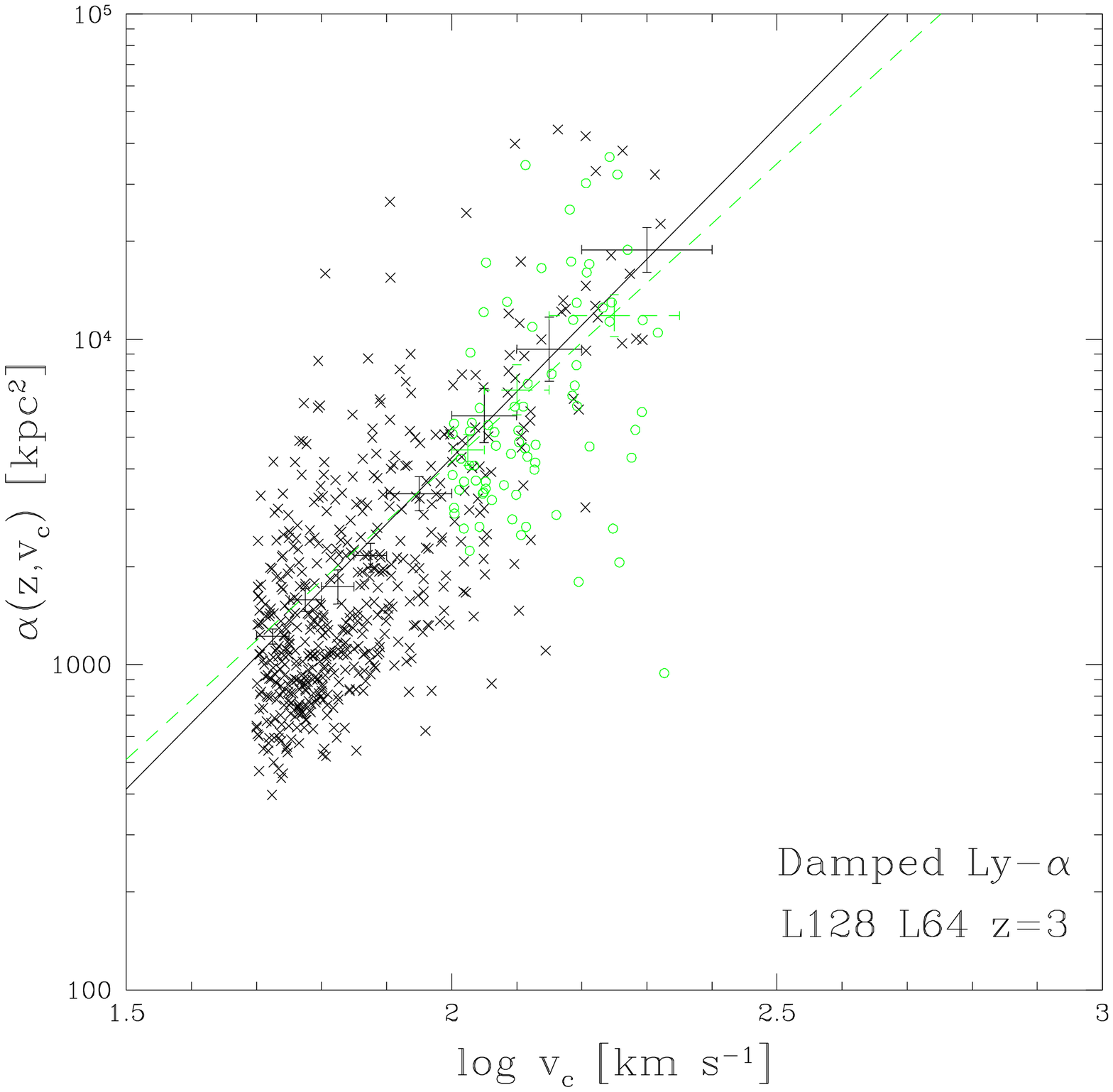}{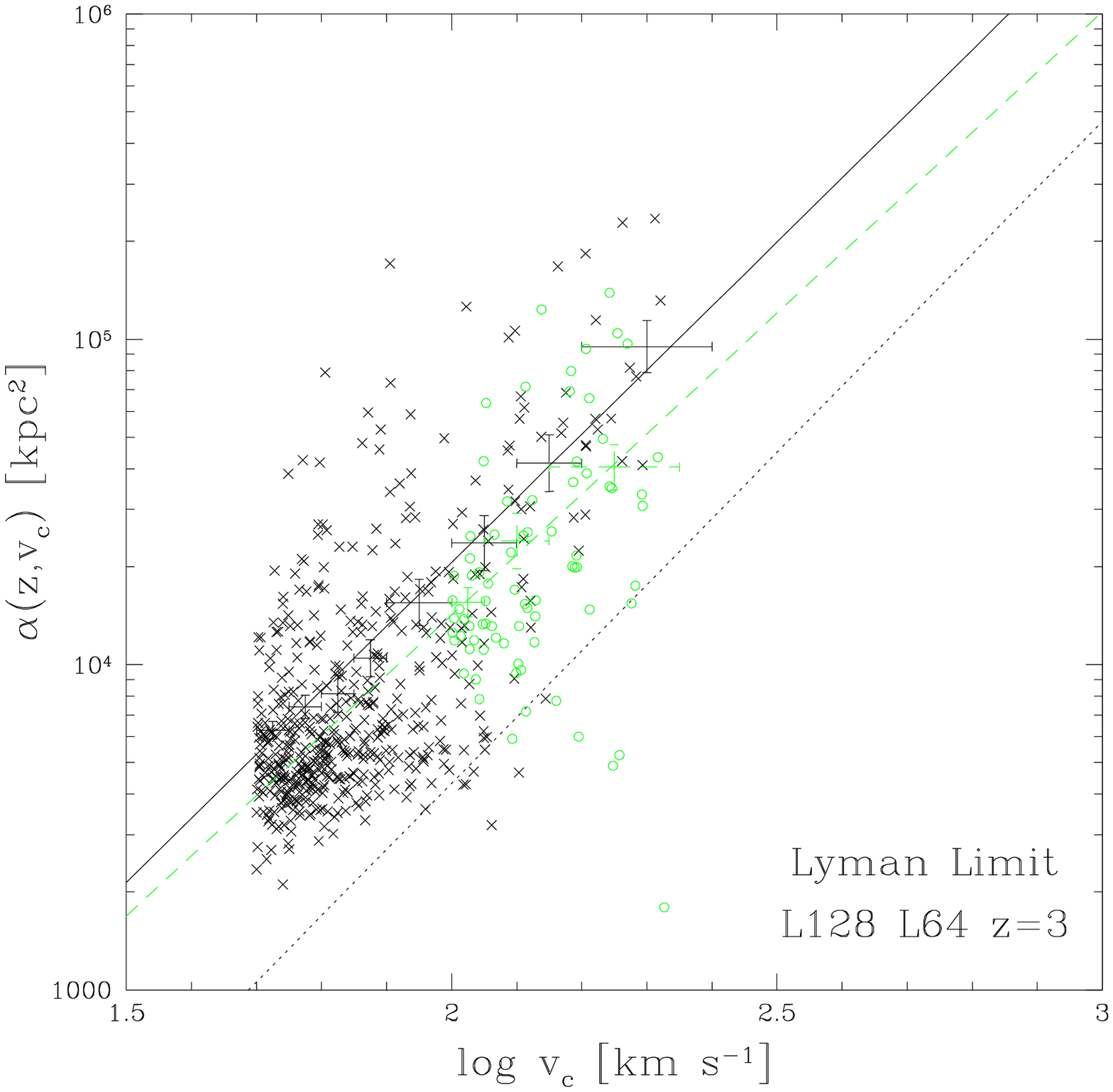}
\caption{
\label{fig:vaL128L64}
Comoving absorbing area (kpc$^2$) vs. circular velocity $v_c$
(\kmpersec) for halos in the L128 (crosses) and L64 (circles) runs at
$z=3$.  As in Figure~\protect\ref{fig:vaL128}, the left panel shows
\dla\ absorption and the right \lyl\ absorption.  The horizontal
components of the error crosses span the bins used to calculate
bootstrap errors on the mean of $\alpha_z(v_c)$, and vertical
error bars show the 1-$\sigma$ limits.  
Details of the error estimation procedure are given in the text;
note that the vertical error bars do not characterize the physical
{\it scatter} in $\alpha(v_c,z)$ but the statistical uncertainty
in the {\it mean value} of $\alpha(v_c,z)$ caused by the finite
number of halos in the bin.
The line is the best fit to the
binned data.  The solid line and error crosses denote the L128 data;
the dashed line and error crosses show L64.  The dotted line on the
right panel corresponds to the best fit line for \dla\ systems (i.e.\
the solid line on the right panel). }
\end{figure}

\begin{figure}
\plottwo{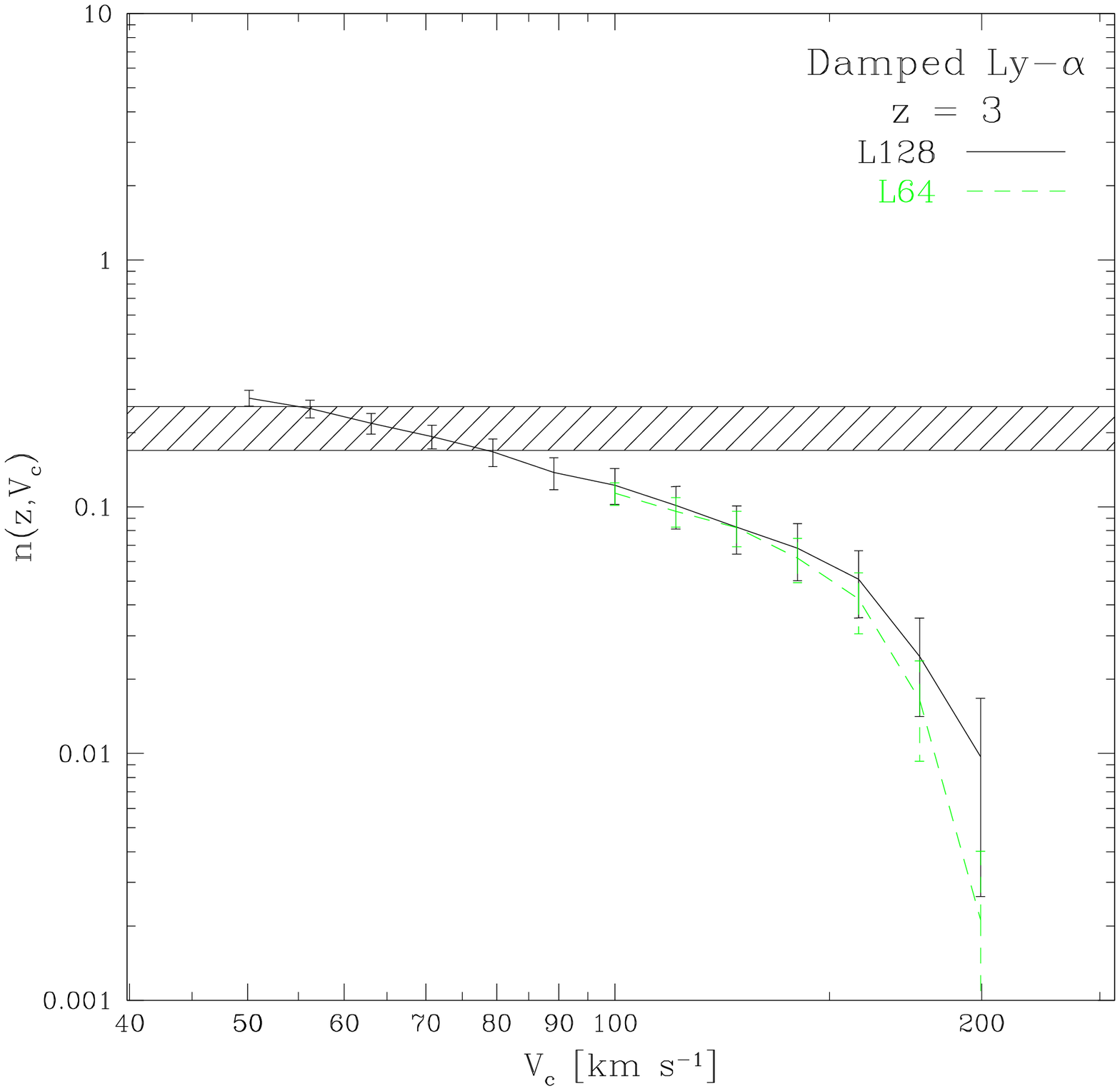}{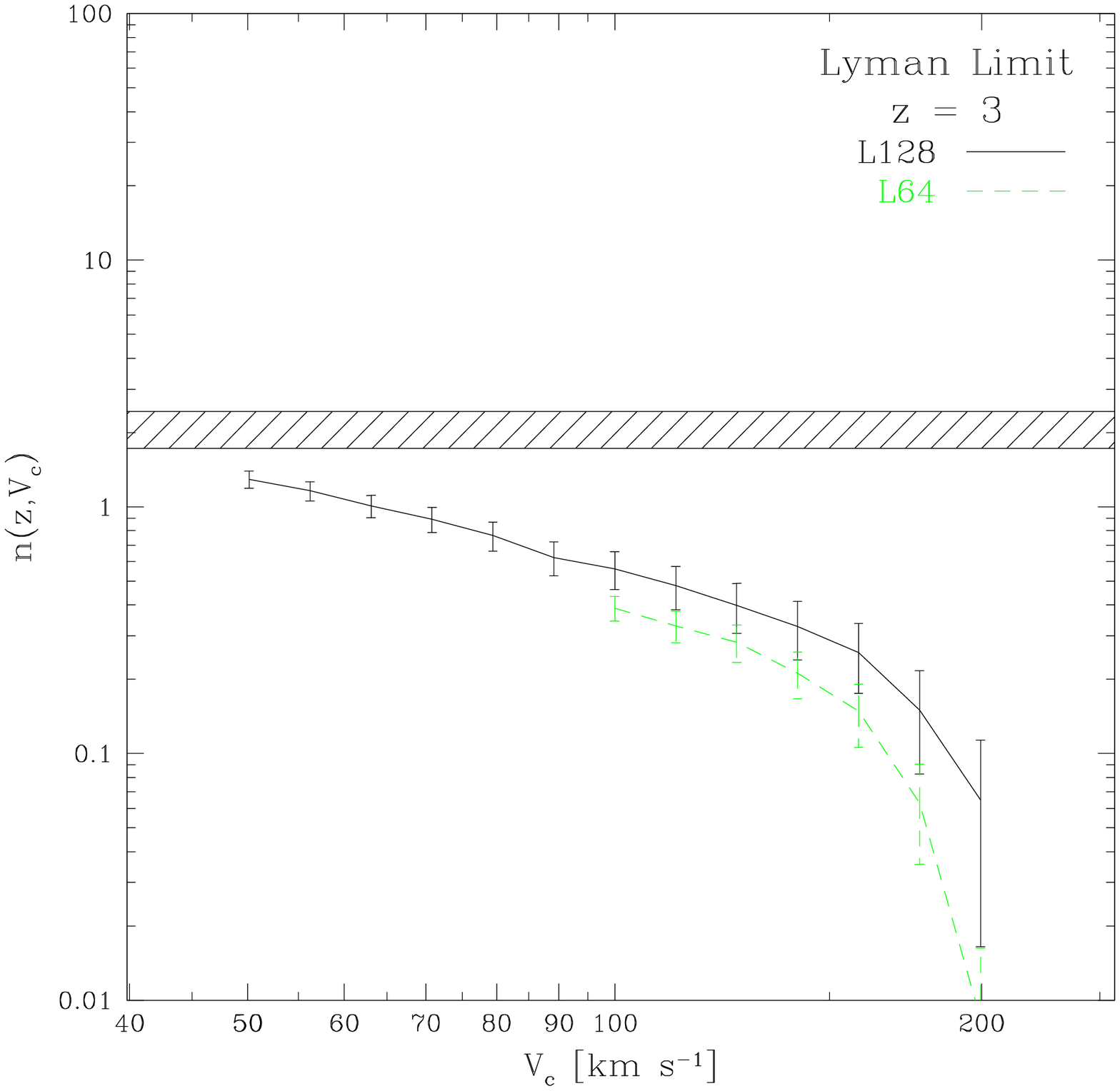}
\caption{
\label{fig:nzplotsim}
Cumulative incidence $n(z,v_c)$ of \dla\ (left panels) and \lyl\
(right panels) systems for the raw simulated data from L128 (solid) and 
L64 (dashed) at $z=3$.  $n(z,v_c)$ is
the number of systems per unit redshift that are located within halos
with circular velocity (at the $\delta=180$ radius) of at least $v_c$.
The cross-hatched region denotes the 1-$\sigma$ observed range of
\dla\ absorption (Storrie-Lombardi \& Wolfe 2000) and \lyl\ absorption
(Storrie-Lombardi et al.\ 1994). }
\end{figure}

\begin{figure}
\vglue-0.65in
\plotone{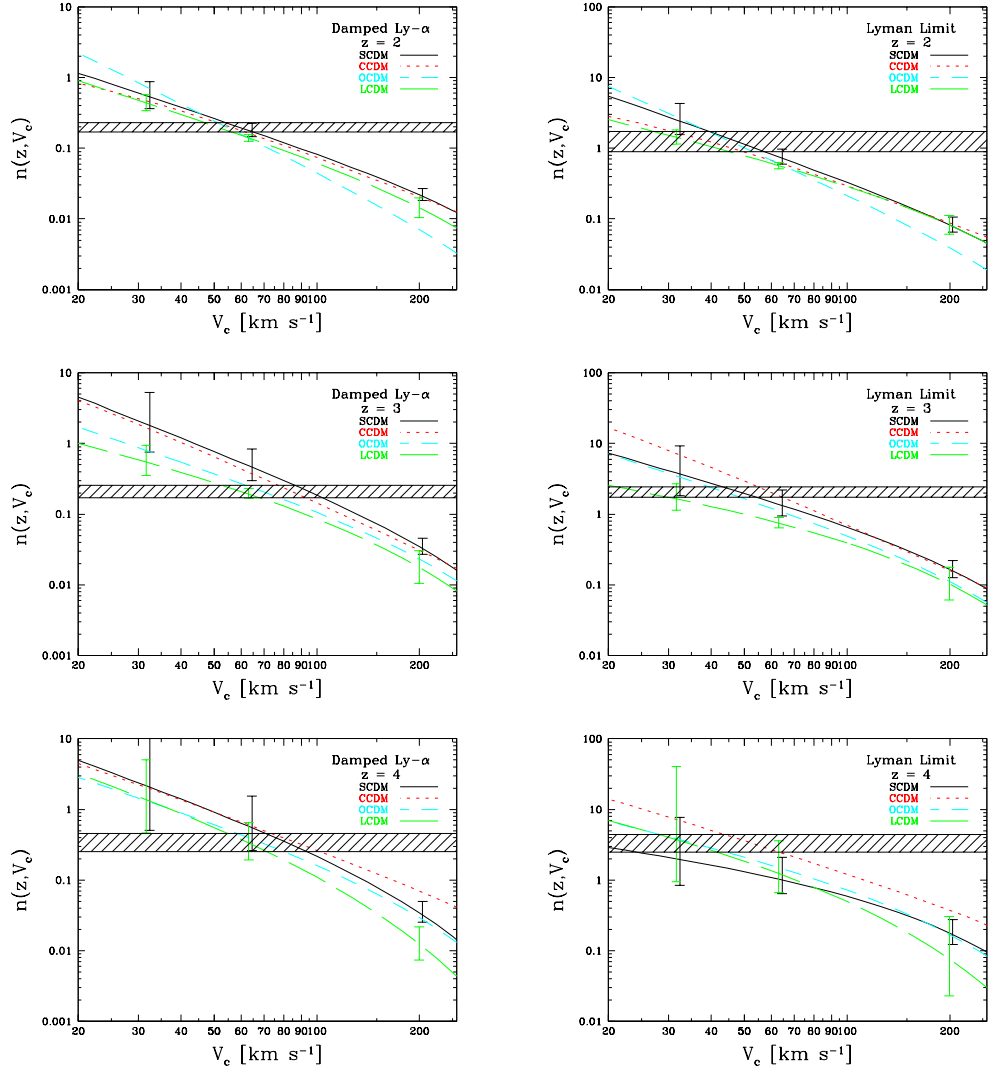}
\vglue-0.26in
\caption{
\label{fig:nzplots}
Cumulative incidence $n(z,v_c)$ of \dla\ (left panels) and \lyl\
(right panels) systems for each model at $z=4,$ 3, 2.  $n(z,v_c)$ is
the number of systems per unit redshift that are located within halos
with circular velocity (at the $\delta=180$ radius) of at least $v_c$.
Representative error bars are plotted for the LCDM and SCDM models, with a
horizontal offset of 0.01 dex applied to SCDM for clarity.  
The cross-hatched region denotes the 1-$\sigma$ observed range of
\dla\ absorption (Storrie-Lombardi \& Wolfe 2000) and \lyl\ absorption
(Storrie-Lombardi et al.\ 1994). }
\end{figure}

Figure~\ref{fig:nzplots} shows the results of this exercise.  The
curves show the cumulative incidence $n(z,v_c)$ as a function of $v_c$
for redshifts $z=(2,3,4)$.  The error bars for SCDM and LCDM are shown
at three representative locations and show the $1-\sigma$ error region
resulting from the bootstrap uncertainty in the fits for $\alpha_{\rm
PL}(z,v_c)$.  It comforting to note that the $n(z,v_c)$ curve for
LCDM, which is closest to the cosmology of L64 and L128, does indeed
mirror the simulated $n(z,v_c)$ shown in Figure~\ref{fig:nzplotsim}.
The cross-hatched region of Figure~\ref{fig:nzplots} denotes the
$1-\sigma$ range allowed observationally by Storrie-Lombardi \& Wolfe
(2000; \dla) and Storrie-Lombardi \ea\ (1994; \lyl).  To match \dla\
observations, contributions from halos with $v_c
\simgt 60\;\vunits$ are required.  In the \lyl\ case, the minimum halo
circular velocity is somewhat lower, more in the range of $v_{c,min} \sim
40\;\vunits$, although given the results from
Figure~\ref{fig:vaL128L64}, the estimate of $n(z,v_c)$ may be
depressed by $\sim 25\%$.  Raising $n(z,v_c)$ by this amount would actually
bring the minimum \lyl-harboring halo mass in line with the \dla\
estimate for most models.  On the other hand, it is possible that
\lyl\ absorbers could reside in halos of lower mass than {\dla}s.

The results show that although halos in models like LCDM generally
have higher absorption cross sections, the increased number density of
halos in the critical models tends to give them more total absorption
than the subcritical models.  Although their $\sigma_8$ values are close to that
of SCDM (Table~\ref{tab:params}), and the growth factor reduction
from $z=0$ to $z=2-4$ is smaller in subcritical models, OCDM and LCDM
also have redder power spectra and thus less power on these relatively
small scales.  The difference in power spectrum shape is especially
important at the low-mass end of the mass function.

\begin{table}
\begin{tabular}{rrrrrrr}
 \tableline\tableline
 \multicolumn{1}{c}{ }& \multicolumn{6}{c}{Damped \lya}\\ \cline{2-7}
 \multicolumn{1}{c}{ }& \multicolumn{2}{c}{$z=2$}&
 \multicolumn{2}{c}{$z=3$}&  \multicolumn{2}{c}{$z=4$} \\
\multicolumn{1}{r}{Model}&
\multicolumn{1}{c}{B}& \multicolumn{1}{c}{log A}&
\multicolumn{1}{c}{B}& \multicolumn{1}{c}{log A}&
\multicolumn{1}{c}{B}& \multicolumn{1}{c}{log A} \\ \tableline
SCDM&	1.196& 0.722&  0.904& 1.777&  1.049& 1.605\\
CCDM&	0.936& 1.232&  0.601& 2.275&  1.036& 1.594\\
OCDM&   0.398& 2.403&  1.207& 1.067&  1.238& 1.254\\
LCDM&   1.247& 0.764&  1.569& 0.325&  1.151& 1.508\\
\tableline\tableline
 \multicolumn{1}{c}{ }& \multicolumn{6}{c}{Lyman Limit}\\ \cline{2-7}
 \multicolumn{1}{c}{ }& \multicolumn{2}{c}{$z=2$}&
 \multicolumn{2}{c}{$z=3$}&  \multicolumn{2}{c}{$z=4$} \\
\multicolumn{1}{r}{Model}&
\multicolumn{1}{c}{B}& \multicolumn{1}{c}{log A}&
\multicolumn{1}{c}{B}& \multicolumn{1}{c}{log A}&
\multicolumn{1}{c}{B}& \multicolumn{1}{c}{log A} \\ \tableline
SCDM&	1.079& 1.585&  1.443& 1.144&  2.213&-0.508 \\
CCDM&	1.045& 1.591&  0.702& 2.737&  1.285& 1.702 \\
OCDM&   0.628& 2.585&  1.277& 1.582&  1.685& 0.928 \\
LCDM&   1.601& 0.649&  2.015& 0.003&  1.675& 1.034 \\
\tableline\tableline

\end{tabular}
\caption{Fitted parameter values for $\alpha_{\rm
PL}(v_c,z)\equiv A v_c^B$.}
\label{tab:fits}
 \end{table}

\section{Conclusions}
\label{sec:conclusions}

We can divide our conclusions into two classes, those that rely only
on the results of our simulations, and those that rely on our
extrapolation of these results via the Press-Schechter method (with
\citealt{jenkins01} mass function) to account for absorption by low mass
halos.  We will treat these two classes of conclusions in turn.

\subsection{Simulation Results}

Our five principal simulations resolve the formation of cold, dense gas
concentrations in halos with $v_c \geq v_{c,res} = 140,$ 160,
$180\;\kms$ (89, 100, $110\;\kms$) at $z=2,$ 3, 4 in the critical
density (subcritical) models.  We employ a further 2 simulations done
in identical cosmologies but with a factor of eight difference in mass
resolution to examine resolution effects.  The lower resolution run, L64,
is the same resolution as the five principal simulations and has
$v_{c,res}$ equivalent to the subcritical (OCDM and LCDM) models.  The
higher resolution run, L128, has $v_{c,res} = 50\;\kms$.  Our clearest 
conclusion is that absorption in halos above the circular velocity thresholds
of the $64^3$ simulations
cannot account for the observed incidence $n(z)$ of \dla\ or \lyl\
absorption or for the amount of cold, collapsed gas, $\Omega_{ccg}$,
in observed \dla\ systems, for any of our five cosmological models
(Figures~\ref{fig:nzloDLA}, \ref{fig:nzloLL}, \ref{fig:omegaccg}).
Higher resolution simulations are unlikely to change this conclusion,
since clumping of the gas on scales below our gravitational softening
length would tend to reduce the absorption cross section rather than
increase it, unless this small scale clumping could produce neutral
condensations in the outskirts of halos where we predict the gas to be
mostly ionized.  The evidence from L128 indicates that there are no
resolution effects above $v_{c,res}$ that affect \dla\ systems,
although estimates of \lyl\ incidence in halos $v_c \geq v_{c.res}$
may be underestimated by 25\% in the principal simulations.

Our models assume $\Omega_b=0.0125h^{-2}$, and a higher baryon
abundance (e.g.\ \citealt{burles98}ab) might increase the predicted
absorption.  We have investigated SCDM models with different
$\Omega_b$ values and find that higher $\Omega_b$ leads to more
absorption per halo as expected, but even a model with
$\Omega_b=0.03125h^{-2}=0.125$ has too little absorption 
at this circular velocity threshold to match the
observations.  We will report further results from this study in a
future paper.

If any of these cosmological models is correct, then a substantial
fraction of high redshift \dla\ absorption must arise in halos
with $v_c \la 100-150\;\kms$.  This conclusion appears consistent
with the imaging of DLA fields, which often reveals no large, bright
galaxies near the line of sight \citep{fontana96,lebrun97,moller98,RT98,TR00}.  However, it implies that the
asymmetric metal-line profiles found by Prochaska \& Wolfe (1997, 1998)
must be interpreted as a signature of non-equilibrium dynamics
\citep{HS98} rather than smooth rotation.

We find a clear and intuitively sensible relationship between high HI
column density absorption and the proximity to galaxies
(Figure~\ref{fig:impacts}).  Damped
systems typically lie within 10-15 kpc of the center of a host galaxy at
$2 \leq z \leq 4$, while lower column densities near the Lyman limit regime
typically occur farther from the host galaxy.  All \dla\ and
\lyl\ absorption in our simulation occurs within collapsed dark
matter halos.  If it were to occur outside halos in the actual
Universe, it would have to be on size scales smaller than we resolve.

The stellar mass in our simulation is generally a steep function of
time in the redshift range $2<z<4$, corresponding to a power law in
$z$ (Figure~\ref{fig:omegaccg}).
The mass in cold collapsed gas, however, remains relatively
fixed, indicating that the rate at which gas is converted into
stars is roughly equal to the rate at which new gas cools out
of ionized halos and condenses into galaxies.
This result is expected if the star formation rate is an increasing
function of gas density, as it is in our numerical formulation (KWH).

In the first 3-d hydrodynamic study of high column density absorption,
KWHM found that the predictions of $n(z)$ from their simulations of
the SCDM model fell a factor of two short of the observed \dla\
abundance but a factor of ten short of the observed \lyl\ abundance.
They speculated that the \dla\ shortfall could be made up by
absorption in lower mass halos but that the \lyl\ shortfall might
imply a distinct physical mechanism for the formation of \lyl\
systems, such as thermal instability on mass scales far below the
simulation's resolution limits \citep{mom96}.  It appears, however,
that for the resolution of the KWHM runs and the principal simulations
presented here, \lyl\ absorption at simulation scales may not have
converged.  If the resolution is increased by a factor of eight (as in the
L128 run), \lyl\ absorption in halos over the same range in mass
increases by 33\%.  This suggests the possibility that standard
cosmological models can explain the observed \lyl\ systems with the
physical processes that already occur in these simulations, albeit in
halos somewhat below our current resolution limits.  Thus \lyl\ and
\dla\ absorption are closely related rather than physically distinct
phenomena, with \lyl\ absorption arising preferentially at larger
galactocentric distances and in less massive halos.  We find no
evidence in the simulations of \lyl\ absorption outside of galaxy dark
matter halos.

\subsection{Absorption in Low Mass Halos}

In our simulations, the halo absorption cross sections $\alpha(v_c,z)$
are determined by complex and competing physical processes.  If we
consider only halos that contain a single gas concentration, then the
absorption cross section can actually decrease slightly with
increasing circular velocity (solid points in
Figure~\ref{fig:vaL128L64}).  However, more massive halos are more
likely to contain multiple gas concentrations, with the net effect
that $\alpha(v_c,z)$ increases with increasing $v_c$.  At $z=2$, the
model with the weakest mass fluctuations (LCDM) tends to have high
$\alpha(v_c)$ (Figure~\ref{fig:nzplots}).  This fact, and the trend
for single-absorber halos, imply that \dla\ and \lyl\ absorption cross
sections are substantially affected by non-equilibrium dynamics:
absorbers get smaller if they have time to cool and condense in a
quiescent dark matter potential well.  Although others have argued
that this behavior may be a numerical artifact (\cf \citealt{maller00}),
the agreement between our L64 and L128 runs and the appearance of
the same trend in L128 suggests that it is not.
Consequently, we suspect that this
non-equilibrium behavior is a real feature of \dla\ and \lyl\ systems,
possibly the geometric counterpart to the complex kinematic behavior
found by Haehnelt et al.\ (1998).  The physical complexity of
$\alpha(v_c,z)$ implies that an accurate fully analytic description of
high column density absorption in CDM models will be difficult to
achieve.  Even the simple expectation that more small scale power
produces a higher incidence of \dla\ and \lyl\ absorption does not
always hold.

In the L64 and L128 runs, for which \dla\ results agree well in the mass
regime of overlap, the mean cross-section for \dla\ absorption is
$\alpha \approx \pi (0.29 R_{vir})^2$, much larger than the simple
estimate $\alpha \sim \pi(0.1R_{vir})^2$ based on collapse of the
baryons to a centrifugally supported disk.  For \lyl\ absorption,
where we find that absorption in equal-mass halos is 25\% lower in L64
than in L128 (which has a factor of eight finer mass resolution), the
cross sections in L128 are described by $\alpha \approx \pi (0.63
R_{vir})^2$.

To estimate the amount of absorption in halos below the
resolution limits of our simulations, we adopted a procedure similar
to that of GKHW, using the numerical results to calibrate
$\alpha(v_c,z)$ and the Jenkins et al.\ (2001) mass function to compute the
halo abundance.  However, relative to GKHW we employed a much more
conservative estimate of $v_{c,res}$ and an improved error estimation
procedure based on bootstrap analysis instead of Poisson errors.
These changes lead to superior $\alpha(v_c,z)$ fits that generally
increase the predicted amount of absorption in halos with $v_c <
v_{c,res}$.  Our new results for $n(z)$ in the SCDM model supersede
those of GKHW, since our new procedures are certainly an improvement,
and our results for $n(z)$ in other models supersede those of GKWH,
since in addition to these technical improvements we now have
numerical simulations of these other models to constrain
$\alpha(v_c,z)$ for $v_c \geq v_{c,res}$.  The bootstrap procedure
yields believable statistical uncertainties in the $n(z,v_c)$
predictions.

Taking our results and error estimates at face value, we find that
four of the cosmological models that we consider are compatible with
observational estimates of the incidence of \dla\ and \lyl\ absorption
at $z=2,$ 3, and 4.  What hinders us in better quantifying the total
incidence in the Universe is our uncertainly in the estimate of the
$v_{c,min}$, the circular velocity at which halos cease to harbor high
column density systems.  Previous studies (QKE;
\citealt{thoul96}) have found this cutoff to be approximately 40
\kmpersec.  However, due to the extreme number density of halos at
this mass, a slight uncertainty in this cutoff results in huge
uncertainties in estimating the total incidence in \dla\ and \lyl\
systems.  Instead we determine, for each cosmology, the value of
$v_{c,min}$ that best matches $n(z)$ observations.  
Reproducing the data of \citet{slw00} and \citet{SL94} requires
$v_{c,min} \sim 60\;\vunits$ for \dla\ systems and $v_{c,min} \sim
40\;\vunits$ for \lyl\ systems, with some dependence on cosmology
and redshift (see Fig.~\ref{fig:nzplots}).

Since the \dla\ values of $v_{c,min}$ are above the expected threshold
caused by photoionization, there is some risk that all of these 
models would predict {\it too much} \dla\ absorption in simulations
that fully resolved the population of absorbing systems.
A model with somewhat less small scale power, such as the lower 
amplitude LCDM model favored by recent Ly$\alpha$ forest studies
\citep{mcdonald00,croft01}, might fare better in this regard,
perhaps matching the observed \dla\ abundance with a $v_{c,min}$
closer to the expected photoionization value.  
We are unable to make predictions for total absorption in the TCDM
model with our current simulations because the paucity of structure
above our resolution threshold makes our extrapolation procedure
unreliable.

Our current simulations provide a number of insights into the physics
of \dla\ and \lyl\ absorption in halos with $v_c \ga 100\;\kms$.
Unfortunately, they also imply that robust numerical predictions
of the incidence of high-redshift \dla\ and \lyl\ absorption will
require simulations that resolve gas dynamics and cooling in halos with
$v_c \sim 30-100\;\kms$, where our analytic modeling predicts a
large fraction of the high column density absorption to occur.
Simulations that resolve such halos exist (e.g.\ QKE; \citealt{navarro97}),
but they do not yet model large enough volumes to predict statistical
quantities like $n(z)$.  Achieving the necessary combination of resolution
and volume is challenging but within reach of current computational
techniques.  Simulations that meet these requirements will also
teach us a great deal about the internal structure of more massive
\dla\ and \lyl\ systems and about the connection between these systems
and the population of high redshift galaxies.

\acknowledgments

We thank Eric Linder for useful discussions. This work was supported
by NASA Astrophysical Theory Grants NAG5-3922, NAG5-3820, and
NAG5-3111, by NASA Long-Term Space Astrophysics Grant NAG5-3525, and
by the NSF under grants ASC93-18185, ACI96-19019, and AST-9802568.
Gardner was supported under NASA Grant NGT5-50078 and NSF Award
DGE-0074228 for the duration of this work.  The simulations were
performed at the San Diego Supercomputer Center.

\vfill\eject

\end{document}